\documentclass{article}
\usepackage{amsmath,amsthm,amssymb}
\usepackage[a4paper]{geometry}
\usepackage{graphicx}
\usepackage[textwidth=8em,textsize=normalsize]{todonotes}
\usepackage{natbib}
\usepackage{lipsum}
\usepackage{float}
\usepackage{subcaption}
\usepackage{dcolumn}
\usepackage{url}
\usepackage{xcolor}
\usepackage{multirow}
\usepackage[linesnumbered,ruled]{algorithm2e}

\theoremstyle{plain}

\theoremstyle{definition}

\DeclareMathAlphabet{\pazocal}{OMS}{zplm}{m}{n}

\newcommand{\mtx}[1]{\mathbf{#1}}
\newcommand{\vct}[1]{\mathbf{#1}}

\def \mA {\mtx{A}}
\def \mC {\mtx{C}}

\def \mI {\mtx{I}}

\def \mM {\mtx{M}}
\def \mP {\mtx{P}}
\def \mQ {\mtx{Q}}
\def \mR {\mtx{R}}

\def \mW {\mtx{W}}
\def \mX {\mtx{X}}

\def \mbeta {\boldsymbol{\beta}}
\def \mrho {\boldsymbol{\rho}}
\def \mdelta {\boldsymbol{\delta}}
\def \mepsilon {\boldsymbol{\varepsilon}}
\def \meta {\boldsymbol{\eta}}
\def \mmu {\boldsymbol{\mu}}

\def \mgamma {\boldsymbol{\gamma}}

\def \mSigma {\boldsymbol{\Sigma}}
\def \mGamma {\mtx{\Gamma}}

\def \mPhi {\mtx{\Phi}}

\def \zero     {\mathbf{0}}
\def \mzero    {\vct{0}}
\def \mone    {\vct{1}}

\def \mm {\vct{m}}
\def \mr {\vct{r}}
\def \ms {\vct{s}}

\def \mw {\vct{w}}
\def \mx {\vct{x}}
\def \my {\vct{y}}

\def \N {\mathcal{N}}
\def \C {\mathcal{C}}
\def \I {\mathcal{I}}
\def \G {\mathcal{G}}
\def \B {\mathcal{B}}
\def \T {\mathcal{T}}

\usepackage{template_tex}

\newcommand\blfootnote[1]{%
  \begingroup
  \renewcommand\thefootnote{}\footnote{#1}%
  \addtocounter{footnote}{-1}%
  \endgroup
}

\begin{document}

\title{Efficient Fully Bayesian Approach to Brain Activity Mapping with Complex-Valued fMRI Data}
\author{Zhengxin Wang \\
Clemson University
\and
Daniel B. Rowe \\
Marquette University
\and
Xinyi Li \\
Clemson University
\and
D. Andrew Brown \blfootnote{\emph{Address for correspondence}: 
D. Andrew Brown, School of Mathematical and Statistical Sciences, Clemson University, Clemson, SC, USA. Email: ab7@clemson.edu} \\
Clemson University}

\date{} 

\maketitle

\begin{abstract}
Functional magnetic resonance imaging (fMRI) enables indirect detection of brain activity changes via the blood-oxygen-level-dependent (BOLD) signal. Conventional analysis methods mainly rely on the real-valued magnitude of these signals. In contrast, research suggests that analyzing both real and imaginary components of the complex-valued fMRI (cv-fMRI) signal provides a more holistic approach that can increase power to detect neuronal activation. We propose a fully Bayesian model for brain activity mapping with cv-fMRI data. Our model accommodates temporal and spatial dynamics. Additionally, we propose a computationally efficient sampling algorithm, which enhances processing speed through image partitioning. Our approach is shown to be computationally efficient via image partitioning and parallel computation while being competitive with state-of-the-art methods. We support these claims with both simulated numerical studies and an application to real cv-fMRI data obtained from a finger-tapping experiment.
\end{abstract}

\vspace{9pt}


\section{Introduction}

Functional magnetic resonance imaging (fMRI) is a non-invasive brain imaging technique that records signals generated by changes in blood oxygenation levels associated with neuronal activity. This so-called blood-oxygenation-level-dependent (BOLD) signal thus facilitates indirect monitoring of brain activity over time \citep{Bandettini1992}. During task-based fMRI experiments, subjects experience intermittent stimuli, such as viewing images or finger tapping. As the brain responds to a particular stimulus, neuronal activity in certain regions intensifies, leading to increased oxygen consumption. This metabolic change subsequently increases the BOLD response in that region. These BOLD fluctuations impact local magnetic susceptibility, thereby affecting the resulting fMRI signal \citep{Lindquist2008}. Empirical studies have demonstrated that the expected BOLD response in an activated brain region, in reaction to binary ``boxcar'' stimuli (repeated identical on-off periods), can be accurately modeled by convolving the boxcar 0-1 stimulus variable with a gamma or double-gamma hemodynamic response function (HRF) \citep{Boynton1996, Lindquist2009}.

Signals generated by magnetic resonance imaging machines are complex-valued with both real and imaginary components due to forward and inverse Fourier transformations that occur in the presence of phase imperfections \citep{Brown2014}. However, most fMRI studies for brain activity mapping only analyze the magnitudes of the MR signals, as the phase components are typically discarded as part of preprocessing. To identify active voxels in response to a stimulus, a linear model is commonly used \citep{Friston1994, Lindquist2008}. Specifically, any voxel (volumetric pixel) whose BOLD signal magnitude significantly changes over time in response to the stimulus will be considered an active voxel. The magnitude-only approach carries several limitations. For one, the magnitude-only models typically operate on the assumption of normally distributed errors. However, even when the original real and imaginary components of the data possess such Gaussian errors, the magnitude follows a Ricean distribution that is approximately normal only for large signal-to-noise ratios (SNRs) \citep{Rice1944, Gudbjartsson1995}. Large SNRs are not always present, making the Gaussian assumption less tenable, thereby losing power. Moreover, by discarding phase information, we ignore half of the available data that may contain information about the underlying neurophysiological processes. On the other hand, using complex-valued fMRI (cv-fMRI) data for analysis has shown promising results. By fully incorporating both real and imaginary components, cv-fMRI studies allow for more comprehensive and accurate models with greater power to detect task-related neuronal activity. Such models often handle SNR more appropriately and make full use of the data at hand, thereby yielding potentially more informative insights into brain activity \citep{Rowe2004, Rowe2005a, Rowe2005c, Rowe2005b, Lee2007, Rowe2007, Rowe2009a, Adrian2018, Yu2018}.

To determine task-related brain activation maps from fMRI signals, fully Bayesian approaches stand out due to their ability to flexibly model spatial and temporal correlations. In this paper, we propose a fully Bayesian model for brain activity mapping using single-subject cv-fMRI time series. Specifically, we aim to determine which voxels' fMRI signal magnitudes (assuming constant phase) change significantly in response to a particular task, as well as the amount of the change. An effective Bayesian approach for fMRI data analysis should fully utilize both the real and imaginary parts of the fMRI data, capture spatiotemporal correlations, provide high prediction accuracy, and be computationally efficient. Although previous studies have made progress in some of these areas \citep{Woolrich2004, Smith2007, Musgrove2016, Bezener2018, Yu2023}, no single model has yet achieved all of these goals. 
Our proposed approach uses autoregressive models for the temporal correlations and Gaussian Markov random fields \cite[GMRFs;][]{RueHeld05} to capture spatial associations in the cv-fMRI data. Moreover, we employ image partitioning and parallel computation to facilitate computationally efficient Markov chain Monte Carlo \citep[MCMC;][]{GelfandSmith90} algorithms. 

The remainder of the paper is organized as follows. Section \ref{Model} details our proposed model, outlines the priors and posteriors, and explains our strategy for brain partitioning. We demonstrate estimation and inference in Section \ref{SimStudy}, where we use simulated datasets to test the performance of our model in terms of the determination of brain activity maps. Section \ref{RealStudy} shows the results of implementing our proposed approach on cv-fMRI data obtained from real finger-tapping experiment. Lastly, Section \ref{Conclusion} summarizes our findings, highlights our contributions, and outlines potential work for future research in this domain.


\section{Model}\label{Model}

In this section, we present our model for brain activity mapping with cv-fMRI data, including an equivalent real-valued representation. We also describe the brain parcellation strategy for parallel computation.  We derive the posterior distribution of the parameters of interest, as well as an MCMC algorithm for accessing it.

\subsection{Model Formulation}

FMRI, both real- and complex-valued, are known to exhibit temporal correlations. This can be captured by autoregressive (AR) error structure. Thus, our complex-valued model is based on that proposed by \cite{Lee2007}, with some modifications. For the $v^{\text{th}}$ voxel, $v=1, ..., V$, the measured signal is modeled as
\begin{equation}\label{model_complex}
    \my^v=\mx\beta^v+\mr^v\rho^v+\mepsilon^v,
\end{equation}
where all terms are complex-valued except $\mx$. The term $\my^v \in \mathbb{C}^T$ is the vector of signals at voxel $v$ collected at evenly-spaced time points, where $T$ is the total observed time points, and $\mx \in \mathbb{R}^T$ is the vector of the expected BOLD response associated with a particular stimulus, with $\beta^v \in \mathbb{C}$ the associated regression coefficient. We assume that low-frequency trends in $\my^v$ have been removed by preprocessing, and that both $\my^v$ and $\mx$ are centered. The term $\mr^v \in \mathbb{C}^T$ is the vector of lag-1 prediction errors for the assumed AR(1) model, with $\rho^v \in \mathbb{C}$ the scalar autoregression coefficient. The AR(1) model has been shown to often be sufficient for capturing temporal dynamics in fMRI data \cite{Cox1996}. We suppose that the error term $\mepsilon^v$ follows the standard complex normal distribution, that is, $\mepsilon^v\sim\C\N_T(\mmu^v=\mzero, \mGamma^v=2\sigma_v^2\mI, \mC^v=\mzero)$, where $\C\N_T$ denotes a complex normal distribution of dimension $T$ with mean $\mmu^v$, complex-valued, Hermitian and non-negative definite covariance matrix $\mGamma^v$, and complex-valued symmetric relation matrix $\mC^v$. In the appendix, we provide details similar to those presented by \cite{Rowe2009a} that demonstrate the equivalence between the model of \cite{Lee2007} and the cv-fMRI model proposed by \cite{Rowe2004} with constant phase. 

\cite{Picinbono1996} and \cite{Yu2018} provide an equivalent real-valued representation of model \eqref{model_complex} as
\begin{equation}\label{model_expended}
\underbrace{\begin{pmatrix}
\my_{Re}^v\\
\my_{Im}^v
\end{pmatrix}}_{\my_r^v}
=
\underbrace{\begin{pmatrix}
\mx & \mzero\\
\mzero & \mx
\end{pmatrix}}_{\mX_r}
\underbrace{\begin{pmatrix}
\beta_{Re}^v\\
\beta_{Im}^v
\end{pmatrix}}_{\mbeta_r^v}
+
\underbrace{\begin{pmatrix}
\mr_{Re}^v & -\mr_{Im}^v\\
\mr_{Im}^v & \mr_{Re}^v
\end{pmatrix}}_{\mR_r^v}
\underbrace{\begin{pmatrix}
\rho_{Re}^v\\
\rho_{Im}^v
\end{pmatrix}}_{\mrho_r^v}
+
\underbrace{\begin{pmatrix}
\mepsilon_{Re}^v\\
\mepsilon_{Im}^v
\end{pmatrix}}_{\mepsilon_r^v},
\end{equation}
where all terms are real-valued. Using the symbols in the underbraces, this is more concisely written as
\begin{equation}\label{model_real}
\my_r^v=\mX_r\mbeta_r^v+\mR_r^v\mrho_r^v+\mepsilon_r^v,\qquad \mepsilon_r^v\sim\N_{2T}(\mzero, \mSigma^v),
\end{equation}
where
\begin{equation}\label{model_error1}
\mSigma^v=
\begin{pmatrix}
\mSigma_{Re, Re}^v & \mSigma_{Re, Im}^v\\
\mSigma_{Im, Re}^v & \mSigma_{Im, Im}^v
\end{pmatrix},
\end{equation}
and
\begin{equation}\label{model_error2}
\begin{matrix}
\mSigma_{Re, Re}^v=\frac{1}{2}Re(\mGamma^v+\mC^v)=\sigma_v^2\mI_T, & \mSigma_{Re, Im}^v=\frac{1}{2}Im(-\mGamma^v+\mC^v)=\mzero_T,\\
\mSigma_{Im, Re}^v=\frac{1}{2}Im(\mGamma^v+\mC^v)=\mzero_T, & \mSigma_{Im, Im}^v=\frac{1}{2}Re(\mGamma^v-\mC^v)=\sigma_v^2\mI_T.
\end{matrix}
\end{equation}
Observe that our assumption on the covariance structure here simply means that $\mSigma^v=\sigma_v^2\mI_{2T}$. We assign the voxel- specific variances $\sigma_v^2$ and autoregression coefficient $\mrho_r^v$ Jeffreys prior and uniform prior, respectively. That is,
$p(\sigma_v^2) = 1/\sigma_v^2$ and $p(\mrho_r^v) = 1$, for $v= 1, \ldots, V$.

\subsection{Brain Parcellation and Spatial Priors}
In addition to temporal dependence, fMRI signals also exhibit spatial associations. These spatial dependencies can originate from several sources, including the inherent noise of the data \citep{Kruger2001}, unmodeled neuronal activation \citep{Bianciardi2009}, and preprocessing steps such as spatial normalization \citep{Friston1995}, image reconstruction \citep{Rowe2009b}, and spatial smoothing \citep{Mikl2008}. Hence voxels, as artificial partitions of the human brain, often exhibit behavior similar to that of their neighbors. These spatial dependencies can be modeled by imposing spatial structure in the prior on $\beta^v$ or the hyperparameters in such priors.

\paragraph{Brain parcellation} \cite{Musgrove2016} propose a brain parcellation technique that seeks to identify active voxels within each parcel/partition, and subsequently combines these results to generate a comprehensive whole-brain activity map. The authors partition their brain images into initial parcels of size approximately 500 voxels each. If a parcel is found to be too large or too small, it is broken down into voxels and these voxels are merged into adjacent parcels while ensuring the merged parcels contain less than 1000 voxels each. Alternatively, the partitioning strategy could be based on anatomical atlases such as Brodmann areas \citep{Amunts2000, Tzourio2002}, or based on equal geometric size in the image rather than equal numbers of contained voxels. \cite{Musgrove2016} remark that this method of partitioning induces negligible edge effects, that is, the classification of voxels on the borders of parcels is not strongly affected.

In our study, we partition the two- or three-dimensional fMRI image into $G$ parcels of approximately equal geometric size. We then process each parcel independently using the same model and method, facilitating parallel computation and hence computational efficiency. We find that our parcellation strategy incurs minimal edge effects, echoing the observations of \cite{Musgrove2016}. We discuss the optimal number of parcels and corresponding number of voxels in each parcel in Section~\ref{SimStudy}.

\paragraph{Prior distribution of $\beta^v$} For parcel $g$, $g=1,\hdots,G$, containing $V_g$ voxels, a voxel $v$ ($v=1,\hdots,V_g$) is classified as an active voxel under the stimulus if its regression coefficient of slope $\beta^v=\beta_{Re}^v+i\beta_{Im}^v\neq0$, where $i$ is the imaginary unit. As this is a variable selection problem, we use a spike-and-slab prior \citep{Mitchell1988, Yu2018}:
\begin{equation}\label{prior_beta_complex}
    \beta^v\mid\gamma_v\sim\gamma_v\C\N_1(0, 2\tau^2_g, 0)+(1-\gamma_v)\I_0,
\end{equation}
where $\I_0$ denotes the point mass at 0. The binary indicator $\gamma_v\in\{0, 1\}$ reflects the status of a voxel. Specifically, $\gamma_v=1$ indicates that voxel $v$ is responding to the task, while $\gamma_v=0$ otherwise. We take $\tau^2_g \in \mathbb{R}$ to be constant across all voxels within each parcel. \cite{Yu2018} shows that a real-valued representation of \eqref{prior_beta_complex} is given by:
\begin{equation}\label{prior_beta_real}
    \mbeta_r^v=
    \begin{pmatrix}
    \beta_{Re}^v\\
    \beta_{Im}^v
    \end{pmatrix}
    \mid\gamma_v
    \sim{\N_2}(\zero, ~\gamma_v\tau_g^2\mI).
\end{equation}
The parcel specific variances $\tau_g^2$ are assigned a Jeffreys prior, $p(\tau_g^2) = 1/\tau_g^2, ~g= 1, \ldots, G$.


\paragraph{Spatial prior on $\gamma_v$} To further reduce computational effort and to capture pertinent spatial structure with a low-dimensional representation, we employ the sparse spatial generalized linear mixed model (sSGLMM) prior, as developed by \cite{Hughes2013} and \cite{Musgrove2016}, which is in turn an extension of the the prior proposed by \cite{Reich2006}. Such priors use GMRFs and reduce the dimension by examining the spectra of the associated Markov graphs. For voxel $v$ ($v=1, ..., V_g$) within parcel $g$ ($g=1, ..., G$), we suppose that
\begin{equation}\label{prior_spatial}
    \begin{split}
        \gamma_v\mid\eta_v&\stackrel {iid}\sim{\B}ern\left\{\mPhi(\psi+\eta_v)\right\}, \\
        \eta_v\mid\mdelta_g&\sim\N_1\left(\mm_v'\mdelta_g, 1\right), \\
        \mdelta_g\mid\kappa_g&\sim\N_q\left\{\mzero, (\kappa_g\mM_g{'}\mQ_g\mM_g)^{-1}\right\}, \\
        \kappa_g&\sim{\G}amma\left(a_{\kappa}, b_{\kappa}\right),
    \end{split}
\end{equation}
where $\mPhi(\cdot)$ denotes the CDF of standard normal distribution and $\psi \in \mathbb{R}$ is a fixed tuning parameter. The terms $\mm_v'$, $\mM_g$, and $\mQ_g$ are derived from the adjacency matrix $\mA_g$ of parcel $g$. The adjacency matrix $\mA_g \in \{0, 1\}^{V_g \times V_g}$ is such that $\mA_{g,uv}=1$ if voxels $u$ and $v$ are neighbors in the image, and 0 otherwise, where ``neighbor'' is defined by the user. Typically, voxels that share an edge or a corner are taken to be neighbors. The matrix $\mM_g \in \mathbb{R}^{V_g \times q}$ contains the first $q$ principal eigenvectors of $\mA_g$, typically with $q\ll V_g$. The term $\mm_v'$ is a $1\times{q}$ row vector of ``synthetic spatial predictors'' \citep{Hughes2013} corresponding to the $v^{\text{th}}$ row of $\mM_g$. The matrix $\mQ_g=\text{diag}(\mA_g{\mone}_{V_g})-\mA_g$ is the graph Laplacian. The term $\mdelta_g$ is a $q\times1$ vector of spatial random effects, and $\kappa_g$ is the spatial smoothing parameter.

The design of the prior distribution for binary indicator $\gamma_v$ aims to capture both spatial dependencies and the sparsity of active voxels. This reflects the hypothesis that a voxel is more likely to be active/inactive if their neighboring voxels are also active/inactive \citep{Friston1994, Smith2007}. Furthermore, in the context of simple tasks, only a small percentage of voxels across the entire brain are expected to be active \citep{Rao1996, Epstein1998}. Thus the sSGLMM prior is well-suited to the work and compatible with the parcellation approach. \cite{Hughes2013} remark that $\mM_g$ is capable of capturing smooth patterns of spatial variation at various scales. 

The parameters $\psi$, $q$, $a_{\kappa}$, and $b_{\kappa}$ are fixed {\em a priori} and determined based on several factors. In our simulation studies, we examine various values of $\psi$ to identify the one providing the highest prediction accuracy. For real human datasets, the initial value of $\psi$ is set to $\mPhi^{-1}(0.02)=-2.05$ for all voxels, following the suggestion of \cite{Musgrove2016}. This value can be further adjusted based on the proportion of active voxels detected in previous experiments. We set $q=5$ (when $V_g$ is approximately 200) per \cite{Hughes2013}, indicating that such a reduction is often feasible. We find there is no detectable difference using larger $q$. The shape and scale parameters of the gamma distribution, $a_{\kappa}=\frac{1}{2}$ and $b_{\kappa}=2000$ respectively, are selected to yield a large mean for $\kappa_g$ ($a_{\kappa}b_{\kappa}$=1000). This choice serves to reduce the chances of creating misleading spatial structures in the posterior distribution, mitigating the risk of identifying spurious brain activity patterns that could be attributed to noise or other confounding factors.

\subsection{MCMC algorithm and posterior distributions}

We use Gibbs sampling to obtain the joint and marginal posterior distributions of parameters of interest. The necessary full conditional distributions and derivations are outlined in the appendix. 
The fixed-width approach proposed by \cite{Flegal2008} is used to diagnose convergence. Specifically, we consider the algorithm to have converged if the Monte Carlo standard error (MCSE) of any $\gamma_v$ is less than 0.05. In our numerical studies that follow, we run $10^3$ iterations. We take the means of the sampled parameters (after discarding burn-in iterations) as the point estimates. Active voxels are determined by $\widehat{\gamma}_v>0.8722$ \citep{Smith2007}, and $\widehat{\beta}^v_{Re}$ and $\widehat{\beta}^v_{Im}$ are used to construct the estimated magnitude maps, computed as $\sqrt{(\widehat{\beta}_{Re}^{v})^2+(\widehat{\beta}_{Im}^{v})^2}$.


\section{Simulation studies}\label{SimStudy}
In this section, we simulate three types of two-dimensional complex-valued time series of fMRI signals: data with {\em iid} noise, data with noise following AR(1) temporal dependence, and a more realistic simulated {\em iid} dataset imitating the human brain. We evaluate three models based on their performance in both classification and estimation fidelity. The models under consideration include:
\begin{itemize}
\item The model of \cite{Musgrove2016}, which uses a sSGLMM prior for magnitude-only data and incorporates brain parcellation (denoted as MO-sSGLMM).
\item The model of \cite{Yu2018} for cv-fMRI, which does not incorporate a spatial prior or brain parcellation (denoted as CV-nonSpatial). In this model, the prior for $\gamma_v$ in model \eqref{prior_spatial} is taken to be $\gamma_v\mid\eta_v\stackrel {iid}\sim{\B}ern(\eta_v), \eta_v\sim{\B}eta(1, 1)$.
\item Our proposed model, which uses an sSGLMM prior for complex-valued data and incorporates brain parcellation (denoted as CV-sSGLMM).
\end{itemize}
All three models are fully Bayesian, suitable for autoregressive noise, and leverage Gibbs sampling to approximate their respective posterior distributions. Both MO-sSGLMM and CV-sSGLMM use the best combination of parcel number $G$ and tuning parameter $\psi$ in terms of the prediction accuracy ($G=9$ and $\psi=\Phi^{-1}(0.47)$ for both), and determine the active voxels by thresholding at $\widehat{\gamma}_v>0.8722$. The CV-nonSpatial model uses a threshold of 0.5, as suggested by \cite{Yu2018}.

Following the model comparisons, we concentrate on our proposed CV-sSGLMM model to examine the impacts of the tuning parameter $\psi$, the number of parcels $G$, and the length of time series $T$. Additional results for marginal posterior distributions, time series, and phase are provided in the appendix.

All of the results are generated by running the code on a custom-built desktop computer with an Intel Core i9-9980XE CPU (3.00GHz, 3001 Mhz, 18 cores, 36 logical processors), NVIDIA GeForce RTX 2080 Ti GPU, 64 GB RAM, and operating on Windows 10 Pro.

\subsection{Simulated datasets with IID noise and AR(1) noise}\label{subsec:SimulatedData}
\begin{figure}
    \begin{center}
        \includegraphics[width=\textwidth]{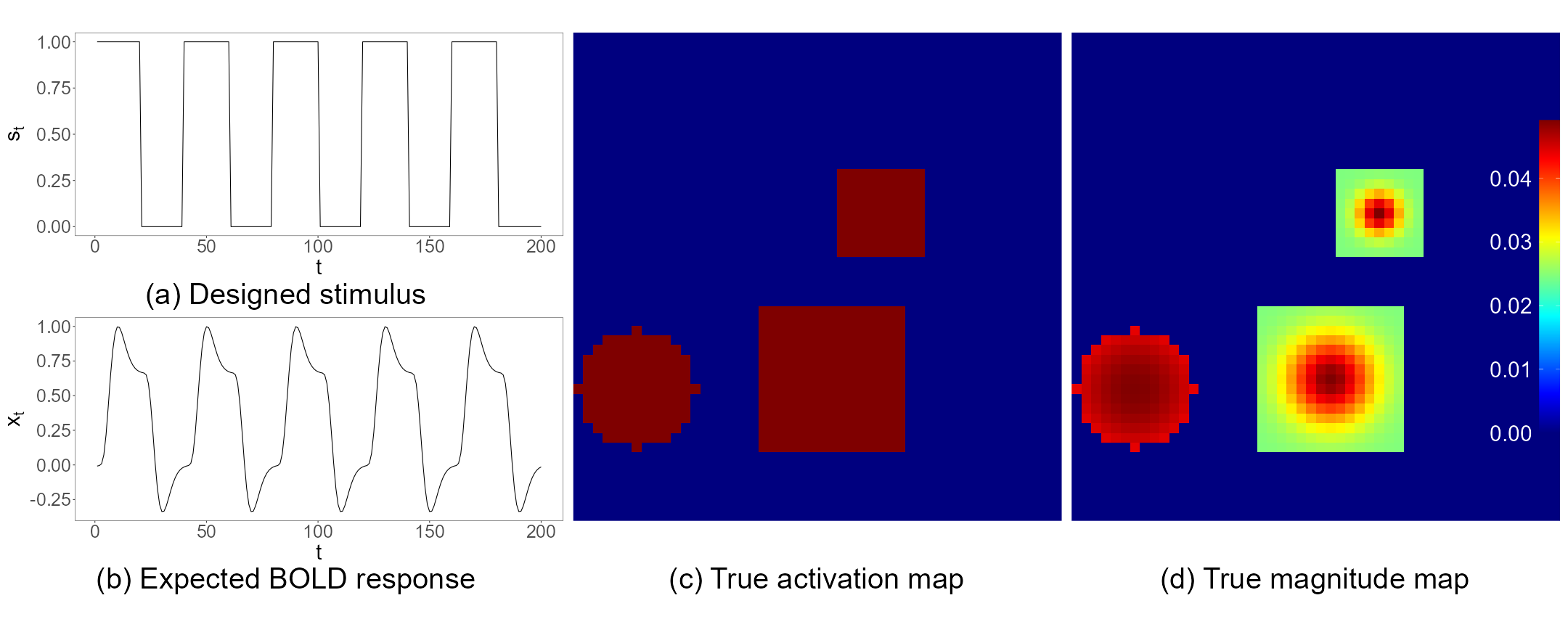}
    \end{center}
    \caption{(a) Designed stimulus; (b) Expected BOLD response; (c) True activation map; (d) True magnitude map.}
    \label{fig:True_maps}
\end{figure}

We discuss how we generate the true maps and simulate fMRI signals here, followed by the results.
\paragraph{Designed stimulus, expected BOLD response, and true activation/magnitude map} We use the same pattern of stimulus as simulated by \cite{Yu2018}. The designed stimulus is a binary signal $\ms$ consisting of five epochs, each with a duration of 40 time points, resulting in a total of $T=200$ time points. Within each epoch, the stimulus is turned on and off for an equal duration of 20 time points. The expected BOLD response, denoted as $\mx$, is generated by convolving the stimulus signal with a double-gamma HRF. Both the designed stimulus and expected BOLD response, depicted in Figures~\ref{fig:True_maps}a and \ref{fig:True_maps}b, are shared for all simulated datasets.

To simulate 100 replicates on a $50\times 50$ panel, we use the \texttt{specifyregion} function in the \texttt{neuRosim} library \citep{Welvaert2011} in \texttt{R} \citep{Rstats}. Each map features three non-overlapping active regions with varying characteristics such as centers, shapes, radii, and decay rates as shown in Table~\ref{tab:map parameters}. The central voxel of an active region has a magnitude of one, while the magnitudes of the surrounding active voxels decrease based on their distance to the center and the decay rate $\varrho$. These magnitudes are further scaled by a multiplier of 0.04909 (which determines to the contrast-to-noise ratio via Eq. \eqref{sim_NoAR}), yielding a range of 0 to 0.04909. Examples of the true activation map and true magnitude map are shown in Figures~\ref{fig:True_maps}c and \ref{fig:True_maps}d.

\begin{table}
    \caption{Characteristics of true maps.}
    \begin{center}
        \begin{tabular}{c|c|c|c|c}
        Map size & Number of active regions & Radius & Shape & Decay rate $(\varrho)$\\\hline
        50$\times$50 & 3 & 2 to 6 & sphere or cube & 0 to 0.3 \\
        \end{tabular}
    \end{center}
    \label{tab:map parameters}
\end{table}

\paragraph{Simulating fMRI signals with non-AR noise and AR(1) noise} We simulate 100 datasets with {\em iid} noise using the expected BOLD response and each true magnitude map for CV-nonSpatial and CV-sSGLMM. We then extract the moduli to use with MO-sSGLMM. The cv-fMRI signal of voxel $v$ at time $t$ is simulated by:
\begin{equation}\label{sim_NoAR}
    \begin{split}
        y_{t, Re}^v&=(\beta_0+\beta^v_1x_t){\rm cos}(\theta)+\varepsilon_{t, Re}^v, \quad \varepsilon_{t, Re}^v\sim\N(0, \sigma^2), \\
        y_{t, Im}^v&=(\beta_0+\beta^v_1x_t){\rm sin}(\theta)+\varepsilon_{t, Im}^v, \quad \varepsilon_{t, Im}^v\sim\N(0, \sigma^2),
    \end{split}
\end{equation}
where $x_t$ represents the expected BOLD response from Figure~\ref{fig:True_maps}b at time $t$, and $\beta^v_1$ refers to the true magnitude of voxel $v$ taken from Figure~\ref{fig:True_maps}d. The phase, $\theta$, is set to be the constant $\pi/4$, and $\sigma$ is set to the constant 0.04909. As a result, the maximum contrast-to-noise ratio (CNR) is $\max{\beta_1^v}/\sigma=1$. We determine the intercept $\beta_0$ based on the signal-to-noise ratio (SNR) such that ${\rm SNR}=\beta_0/\sigma=10$, leading to $\beta_0=0.4909$.

Next, we generate 100 datasets with AR(1) noise in a similar manner as Eq. \eqref{sim_NoAR}.  The difference lies in the simulation of error terms, which is done so that
    \begin{equation}\label{sim_AR1_error_iso}
    \begin{pmatrix}
    \varepsilon_{t, Re}^v\\
    \varepsilon_{t, Im}^v
    \end{pmatrix}
    =
    \begin{pmatrix}
    0.2 & -0.9\\
    0.9 & 0.2
    \end{pmatrix}
    \begin{pmatrix}
    \varepsilon_{t-1, Re}^v\\
    \varepsilon_{t-1, Im}^v
    \end{pmatrix}
    +
    \begin{pmatrix}
    \xi_{Re}^v\\
    \xi_{Im}^v
    \end{pmatrix}
    ,\quad
    \begin{pmatrix}
    \xi_{Re}^v\\
    \xi_{Im}^v
    \end{pmatrix}
    \sim
    \N_2\left(\mzero, \sigma^2\mI\right).
\end{equation}
This is a real-valued equivalent of the complex AR(1) error model,
\begin{equation}\label{sim_AR1_error_cp}
    \varepsilon_{t}^v = (0.2+0.9i)\varepsilon_{t-1}^v+\xi_v, \quad \xi_v\sim\C\N_1(0, 2\sigma^2, 0).
\end{equation}

\begin{figure}
    \begin{center}
    \includegraphics[width=\textwidth]{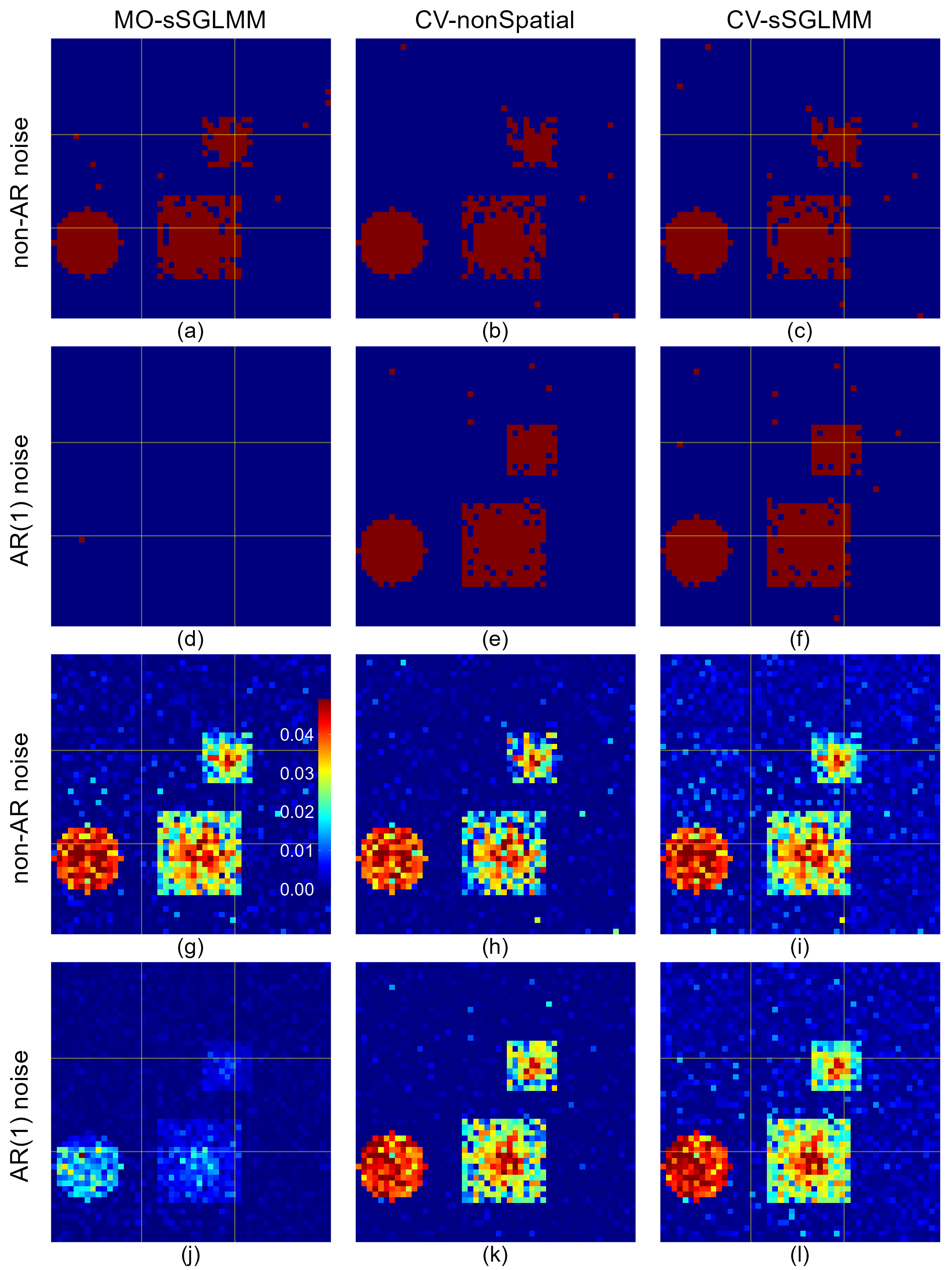}
    \end{center}
    \caption{(a)-(c) are estimated activation maps for a non-AR dataset as produced by the MO-sSGLMM, CV-nonSpatial, and CV-sSGLMM models, respectively. (d)-(f) are estimated activation maps for an AR(1) dataset, as derived from the same models. (g)-(l) are the corresponding estimated magnitude maps.}
    \label{fig:MySim_estimated_maps}
\end{figure}

\begin{figure}
    \begin{center}
    \includegraphics[width=\textwidth]{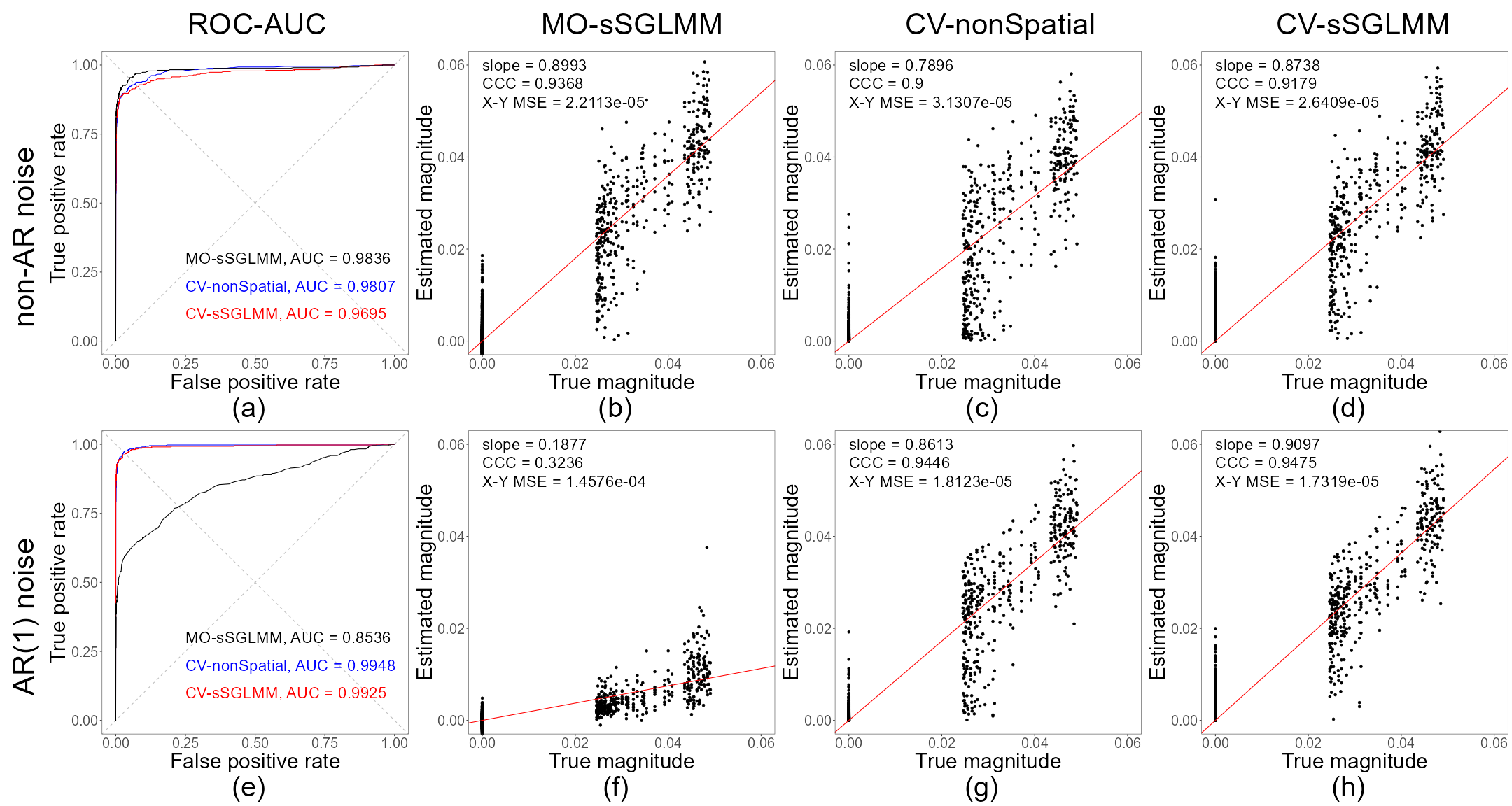}
    \end{center}
    \caption{(a)-(d) are the ROC curves and plots comparing true versus estimated magnitudes for a non-AR dataset. (e)-(h) are analogous plots for an AR(1) dataset.}
    \label{fig:ROC_and_strength}
\end{figure}

\paragraph{Results} Results from our simulations are displayed in Figure~\ref{fig:MySim_estimated_maps}, which depicts the estimated maps for a single dataset. The yellow grid lines correspond to the partitions in cases of brain parcellation. The performance across the three models reveals a consistent trend. All models perform well for the {\em iid} case, while MO-sSGLMM fails to detect any activity in the presence of the AR(1) noise.  This is because the complex-valued AR structure in equation \eqref{sim_AR1_error_cp} cannot be recovered after extracting the moduli of the data. Further quantitative results, such as the receiver operating characteristic area under curve (ROC-AUC), true vs estimated magnitude regression slope, the concordance correlation coefficient (CCC), and true vs estimate pairwise mean square error (X-Y pairwise MSE), are illustrated in Figure~\ref{fig:ROC_and_strength}. These offer a comprehensive performance evaluation in terms of classification and estimation. Figure \ref{fig:ROC_and_strength} shows similar comparative performance as can be gleaned from Figure \ref{fig:MySim_estimated_maps}. All procedures do well in the presence of {\em iid} noise, whereas both complex-valued models considerably outperform the magnitude-only model when the errors are correlated. In each case, we can observe slightly better MSE, CCC, and estimation fidelity (Figure \ref{fig:ROC_and_strength}(b), (c), (d), (f), (g), (h)), but these are small when compared to the outperformance of the complex-valued models versus magnitude only.

Table~\ref{tab:MySim} summarizes the average metrics across 100 {\em iid} noise and 100 AR(1) noise replicated datasets. In the {\em iid} case, the F1-score, slope, CCC, and X-Y MSE clearly favor MO-sSGLMM, followed by our CV-sSGLMM, and CV-nonSpatial ranks last. This demonstrates the proficiency of MO-sSGLMM on datasets where the necessity to capture complex-valued noise dependence is not crucial. The ROC-AUC score of MO-sSGLMM is comparable to that of CV-nonSpatial, and slightly surpasses that of our proposed CV-sSGLMM. 

In the analysis of AR(1) datasets, our proposed CV-sSGLMM shows a clear advantage over the two competitors. Due to MO-sSGLMM's limitations already shown, we focus our comparison here between CV-nonSpatial and CV-sSGLMM. The CV-sSGLMM outperforms CV-nonSpatial across multiple metrics, such as F1-score, slope, CCC, and X-Y MSE. The superior performance of the CV-sSGLMM in terms of both classification and estimation can be attributed to the inclusion of the sSGLMM prior. In addition to our results, the value of using spatial priors to enhance the model's performance on correlated datasets has been demonstrated by \cite{Yu2023}. Perhaps the most notable and favorable performance of our proposed model is in the vastly computational efficiency due to the brain parcellation and parallel computation, 5.39 seconds with CV-sSGLMM versus 42.2 seconds for the CV-nonSpatial. In other words, we obtain results as good or better than current state-of-the-art, but are able to do so 87\% faster.

\begin{table}
    \caption{Summary of average metrics across 100 non-AR and 100 AR(1) datasets produced by the MO-sSGLMM, CV-nonSpatial, and CV-sSGLMM models.}
    \begin{center}
    \renewcommand{\arraystretch}{1.2}
    {\scriptsize
    \setlength{\tabcolsep}{1.7pt}
    \begin{tabular}{c|c|ccccccccc}
    AR type & Mode & Accuracy & Precision & Recall & F1 Score & AUC & Slope & CCC & X-Y MSE & Time (s)\\
    \hline
    \multirow{3}{*}{non-AR} & MO-sSGLMM & \textbf{0.9693} & 0.9440 & \textbf{0.8160} & \textbf{0.8741} & \textbf{0.9774} & \textbf{0.8586} & \textbf{0.9008} & \textbf{2.06e-5} & \textbf{2.4}\\
     & CV-nonSpatial & 0.9540 & \textbf{0.9632} & 0.6687 & 0.7853 & 0.9751 & 0.6771 & 0.8222 & 3.04e-5 & 41.9 \\
     & CV-sSGLMM & 0.9622 & 0.9277 & 0.7742 & 0.8424 & 0.9625 & 0.8186 & 0.8627 & 2.54e-5 & 5.51 \\
    \hline
    \multirow{2}{*}{AR(1)} & CV-nonSpatial & 0.9765 & \textbf{0.9733} & 0.8407 & 0.9012 & \textbf{0.9927} & 0.8040 & 0.9096 & 1.69e-5 & 42.2 \\
     & CV-sSGLMM & \textbf{0.9797} & 0.9381 & \textbf{0.9039} & \textbf{0.9201} & 0.9879 & \textbf{0.8816} & \textbf{0.9145} & \textbf{1.60e-5} & \textbf{5.39} \\
    \end{tabular}
    }
    \end{center}
    \label{tab:MySim}
\end{table}

\begin{figure}
    \begin{center}
    \includegraphics[width=\textwidth]{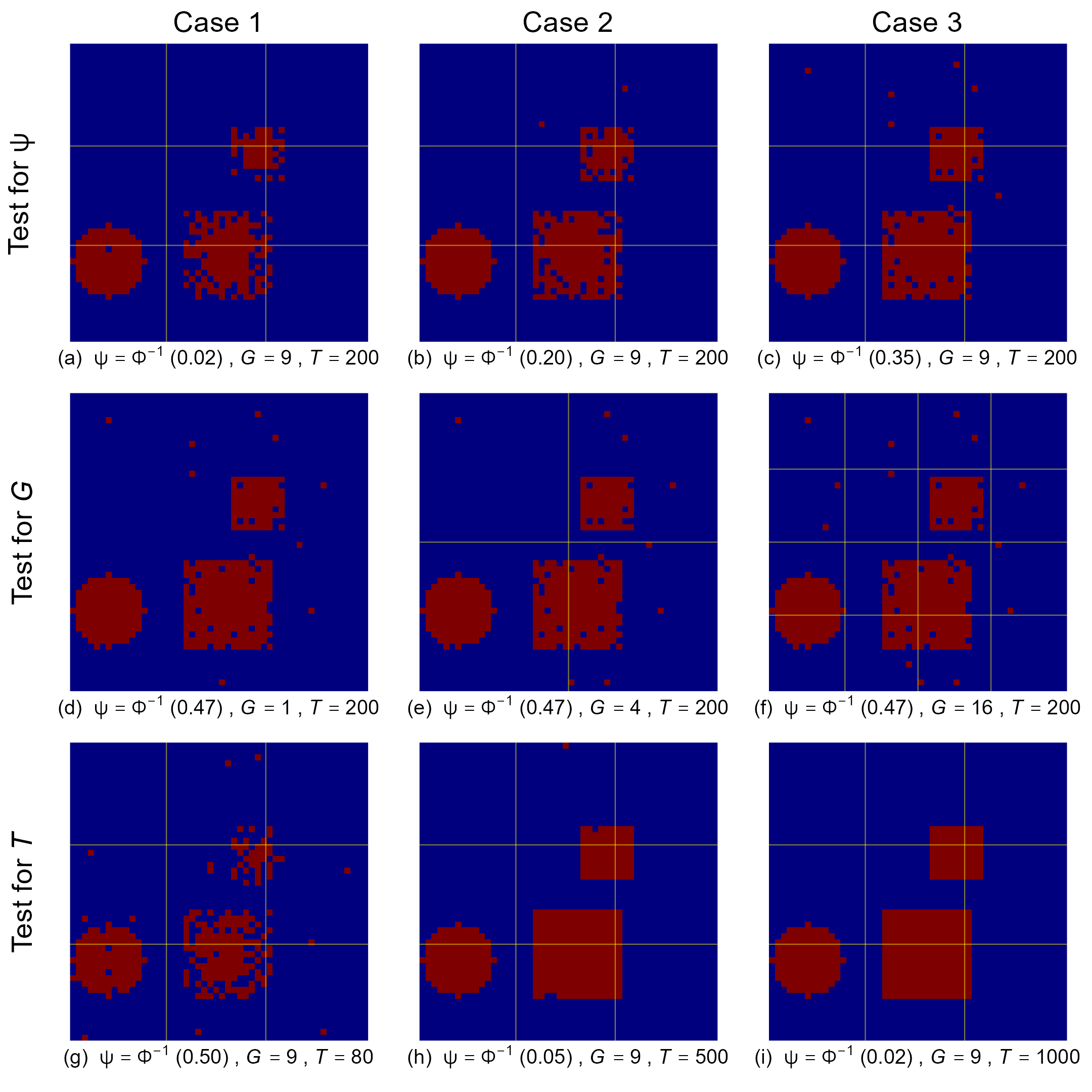}
    \end{center}
    \caption{(a)-(c) are estimated activation maps for an AR(1) dataset as produced by the CV-sSGLMM model using various tuning parameters $\psi$'s. (d)-(f) are estimated activation maps using various parcel numbers $G$'s. (g)-(i) are estimated activation maps derived from datasets with various time lengths $T$'s.}
    \label{fig:Parameter_activation_maps}
\end{figure}

\begin{figure}
    \begin{center}
    \includegraphics[width=\textwidth]{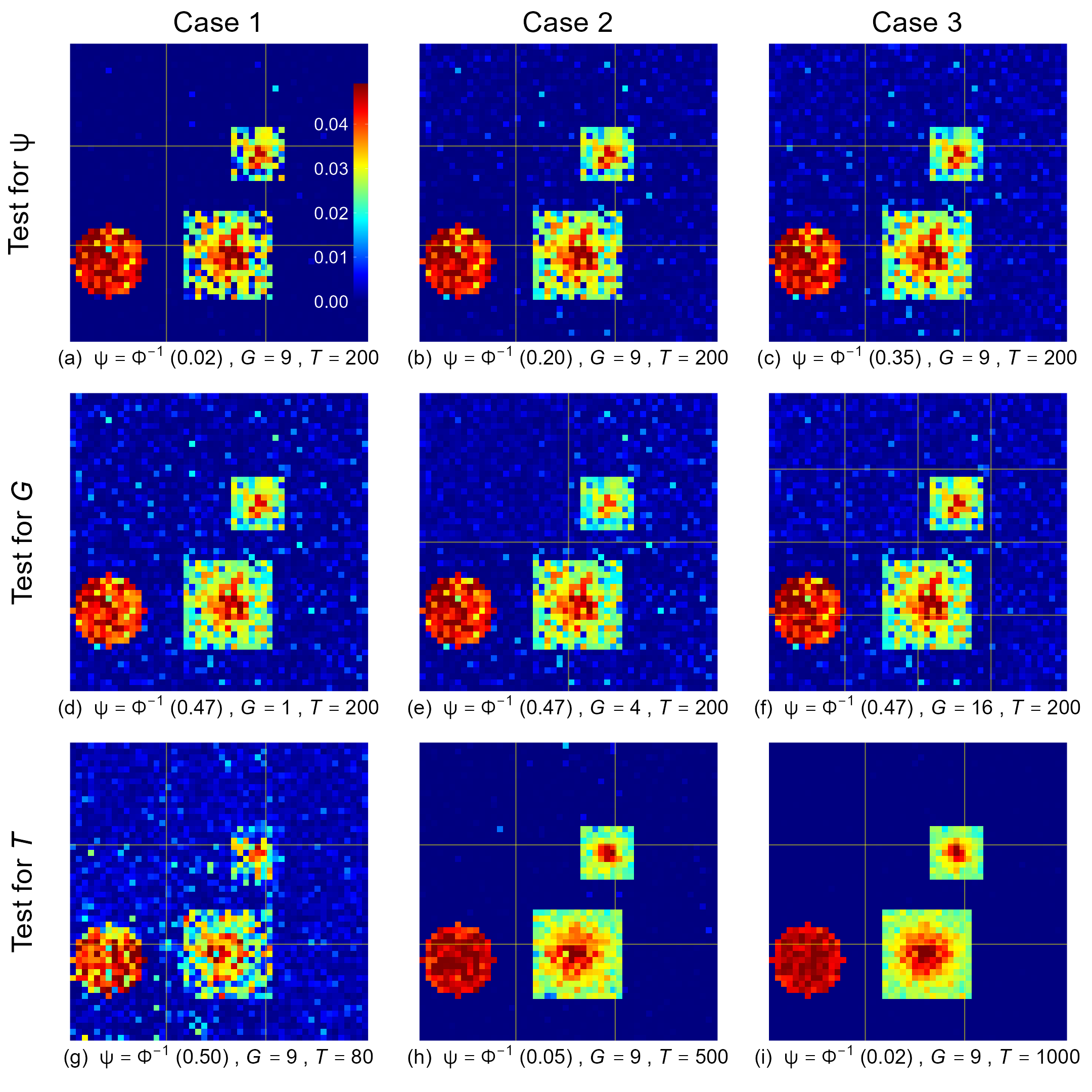}
    \end{center}
    \caption{(a)-(c) are estimated magnitude activation maps for an AR(1) dataset as produced by the CV-sSGLMM model using various tuning parameters $\psi$'s. (d)-(f) are estimated magnitude maps using various parcel numbers $G$'s. (g)-(i) are estimated magnitude maps derived from datasets with various time lengths $T$'s.}
    \label{fig:Parameter_strength_maps}
\end{figure}

\paragraph{Effects of experimental and parameter settings on CV-sSGLMM} The performance of our CV-sSGLMM is determined in part by three choices: the tuning parameter $\psi$, the parcel number $G$, and the time length $T$. Here we assess their influence using the AR(1) data exclusively. For a single dataset, estimated activation maps generated from varying these settings are depicted in Figure~\ref{fig:Parameter_activation_maps}, with their corresponding estimated magnitude maps displayed in Figure~\ref{fig:Parameter_strength_maps}. A summary of average metrics over 100 replicated datasets is shown in Table~\ref{tab:Parameter test}.

Figure~\ref{fig:Parameter_activation_maps}(a)-(c) illustrates the results using $\psi$ values of $\Phi^{-1}(0.02)$, $\Phi^{-1}(0.20)$, $\Phi^{-1}(0.35)$, respectively, which govern the {\em a priori} likelihood of a voxel being determined active. Along with Figure~\ref{fig:MySim_estimated_maps}(f) using $\psi=\Phi^{-1}(0.47)$, we can observe a trade-off in selecting $\psi$: larger values lead to an increase in active voxels and false positives, whereas smaller values result in fewer active voxels and increased false negatives, all of which are as expected. In a simulated scenario, the optimal $\psi$ can be determined by maximizing metrics like prediction accuracy or F1-score. In practical applications, $\psi$ can be tuned to achieve a target percentage of active voxels based on prior experiments, cross-validation, WAIC \citep{Watanabe10}, etc. 

The effects of varying $G=1, 4, 16$ are exhibited in Figure~\ref{fig:Parameter_activation_maps}(d)-(f), respectively. Along with Figure~\ref{fig:MySim_estimated_maps}(f) using $G=9$, we observe negligible edge effects, that is, voxel classifications at parcel borders remain unaffected. Some metrics, such as F1-score, slope, CCC, and X-Y MSE, even exhibit slight improvements through $G=1, 4, 9$. Moreover, the computation time drops significantly as $G$ increases, as expected. These results coincide with the findings of \cite{Musgrove2016}. However, with $G=16$, performance starts decreasing compared to that of using $G=9$ due to insufficient numbers voxels within each parcel. The choice of $G$ and corresponding parcel size $V_g$ can be guided by prior experience or domain-specific knowledge of, e.g., anatomical regions.

Figure~\ref{fig:Parameter_activation_maps}(g)-(i) depicts the impact of varying the time length $T=80, 500, 1000$, respectively. The length of each epoch remains the same as 40 time points so that the number of epochs will change correspondingly. Along with Figure~\ref{fig:MySim_estimated_maps}(f) using $T=200$, we observe improvements in both classification and estimation as $T$ increases. in this case, an accuracy of 100\% is achieved when $T=1000$, and its estimated magnitude map almost perfectly reproduces the truth. It is worth noting that we adopt a relatively low $\psi=\Phi^{-1}(0.02)$ for $T=1000$, suggesting a stringent selection of active voxels. Thus, when an ample number of repeated epochs are available for the stimulus, the signal is strong enough to let us select most of the positive voxels while avoiding false positives. This suggests that choosing a low $\psi$ can enhance discriminative capability.

\begin{table}
    \begin{center}
    \renewcommand{\arraystretch}{1.2}
    {\scriptsize
    \setlength{\tabcolsep}{3pt}
    \begin{tabular}{c|ccccccccc}
    Parameter & Accuracy & Precision & Recall & F1 Score & AUC & Slope & CCC & X-Y MSE & Time (s)\\
    \hline
    $\psi=\Phi^{-1}(0.02)$ & 0.9486 & \textbf{0.9985} & 0.6179 & 0.7585 & 0.9908 & 0.8180 & 0.8924 & 2.16e-5 & 5.39\\
    $\psi=\Phi^{-1}(0.20)$ & 0.9728 & 0.9823 & 0.8123 & 0.8880 & \textbf{0.9878} & \textbf{0.8894} & \textbf{0.9316} & \textbf{1.38e-5} & 5.66 \\
    $\psi=\Phi^{-1}(0.35)$ & \textbf{0.9783} & 0.9628 & \textbf{0.8706} & \textbf{0.9136} & 0.9876 & 0.8893 & 0.9251 & 1.47e-5 & 5.71 \\
    \hline
    $G=1$ & 0.9784 & 0.9220 & 0.9096 & 0.9151 & \textbf{0.9929} & 0.8129 & 0.8818 & 2.09e-5 & 63.37 \\
    $G=4$ & \textbf{0.9796} & \textbf{0.9381} & \textbf{0.9352} & 0.9064 & 0.9908 & 0.8464 & 0.9010 & 1.83e-5 & 12.06 \\
    $G=16$ & 0.9787 & 0.9306 & 0.9045 & \textbf{0.9167} & 0.9874 & \textbf{0.8944} & \textbf{0.9142} & \textbf{1.68e-5} & \textbf{3.74} \\
    \hline
    $T=80$ & 0.9325 & 0.9070 & 0.5449 & 0.6765 & 0.8952 & 0.6961 & 0.7537 & 4.22e-5 & \textbf{3.62} \\
    $T=500$ & 0.9986 & 0.9982 & 0.9919 & 0.9951 & 0.9999 & 0.9749 & 0.9881 & 2.59e-5 & 11.98 \\
    $T=1000$ & \textbf{0.9999} & \textbf{0.9997} & \textbf{1} & \textbf{0.9999} & \textbf{1} & \textbf{0.9889} & \textbf{0.9950} & \textbf{0.11e-5} & 21.17 \\
    \end{tabular}
    }
    \end{center}
    \caption{Summary of average metrics across 100 AR(1) datasets produced by the CV-sSGLMM model using different parameters}
    \label{tab:Parameter test}
\end{table}


\subsection{Realistic simulation}\label{subsec:SimReal}
\begin{figure}
    \begin{center}
    \includegraphics[width=\textwidth]{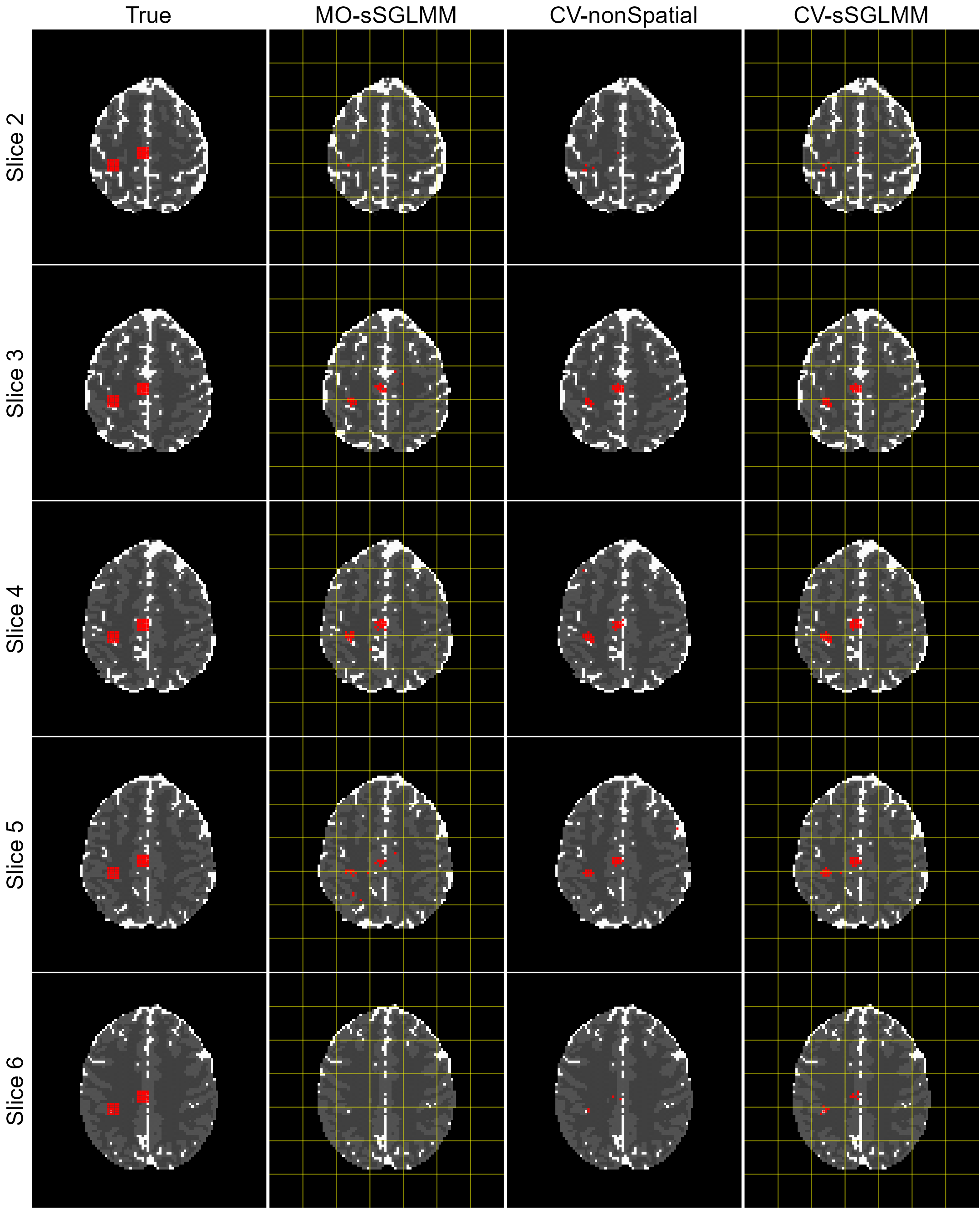}
    \end{center}
    \caption{True and estimated activation maps for a realistic simulation as produced by the MO-sSGLMM, CV-nonSpatial, and CV-sSGLMM models}
    \label{fig:SimReal_activation_maps}
    \end{figure}

\begin{figure}
    \begin{center}
    \includegraphics[width=\textwidth]{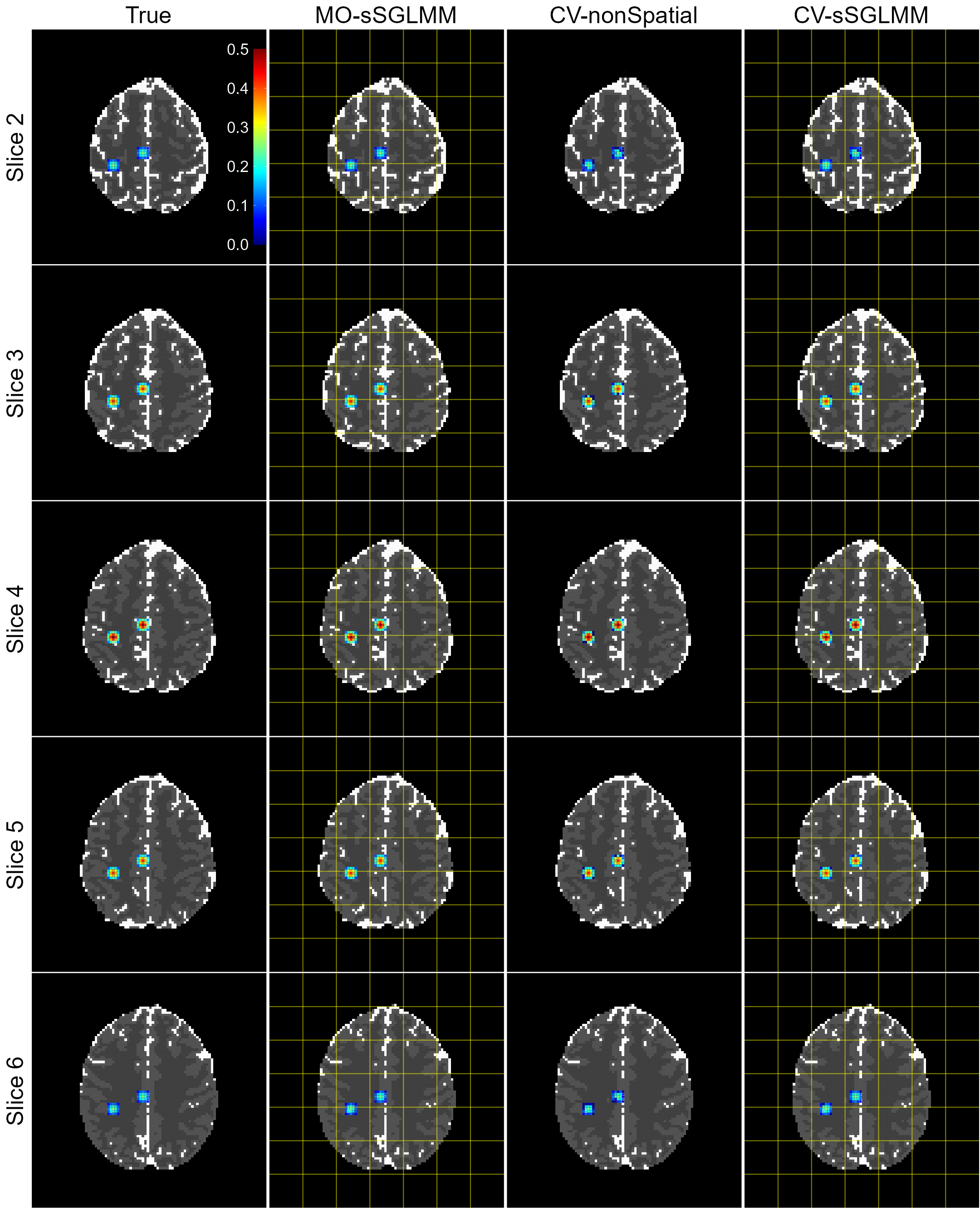}
    \end{center}
    \caption{True and (scaled) estimated magnitude maps for a realistic simulation as produced by the MO-sSGLMM, CV-nonSpatial, and CV-sSGLMM models}
    \label{fig:SimReal_strength_maps}
\end{figure}

Here we simulate a dataset similar that that done by \cite{Yu2018} in which we mimic the environmental conditions of a human brain. The data contain {\em iid} noise. The dataset comprises seven slices, each of size $96\times 96$ voxels, with signals generated across $T=490$ time points. The brain's active regions are two $5\times 5\times 5$ cubes formed by two $5\times 5$ squares within each of slice 2-6. In contrast to the data produced by Eq. \eqref{sim_NoAR}, which exhibits a constant phase, this dataset has a dynamic phase. The cv-fMRI signal for voxel $v$ at time $t$ is thus simulated as
\begin{equation}\label{sim_Real}
    \begin{split}
        y_{t, Re}^v&=(\beta_0+\beta^v_1x_t){\rm cos}(\theta_0+\theta^v_1x_t)+\varepsilon_{t, Re}^v, \quad \varepsilon_{t, Re}^v\sim\N(0, \sigma^2),\\
        y_{t, Im}^v&=(\beta_0+\beta^v_1x_t){\rm sin}(\theta_0+\theta^v_1x_t)+\varepsilon_{t, Im}^v, \quad \varepsilon_{t, Im}^v\sim\N(0, \sigma^2).
    \end{split}
\end{equation}
The slice with the greatest maximum magnitude and phase CNR is slice 4 (Eq. \eqref{sim_Real_CNR}):
\begin{equation}\label{sim_Real_CNR}
    \begin{split}
        \text{CNR}_\text{Mag}&=(\max{\beta_1^v})/\sigma=0.5/1, \\
        \text{CNR}_\text{Ph}&=(\max{\theta_1^v})/{\text{SNR}_\text{Mag}}=(\pi/120)/25.
    \end{split}
\end{equation}
Activation then decreases from slice 4 to slices 3 and 5 and is weakest in slices 2 and 6. Slices 1 and 7 exhibit no activation. It's important to note that, with dynamic phase, the model from \cite{Lee2007} is not equivalent to that from \cite{Rowe2005c} as indicated in \cite{Rowe2009a}. This discrepancy suggests the proposed model is under model misspecification in this scenario. However, as both $\beta_{Re}^v$ and $\beta_{Im}^v$ in model \eqref{model_expended} include magnitude and phase information, and given that prior studies \citep{Yu2018, Yu2023} have used the \cite{Lee2007}-based model to process this dataset, we deem it worthwhile to test our model on these data. We set $G=49$ and a threshold of 0.8722 for both MO-sSGLMM and CV-sSGLMM, with $\psi$ set to $\Phi^{-1}(0.50)$ and $\Phi^{-1}(0.11)$, respectively. For CV-nonSpatial, the threshold is set to 0.5, again following the advice of \cite{Yu2018}. Activation maps are presented in Figure~\ref{fig:SimReal_activation_maps}. We indeed observe that our model tends to overestimate the magnitude.  Since the magnitudes are overestimated, we scale the estimated magnitude to the range of true magnitude in the corresponding slice. True and (scaled) estimated magnitude maps are displayed in Figure~\ref{fig:SimReal_strength_maps}. 

Further numerical results, displayed in Table~\ref{tab:SimReal}, show a pattern of the CV-sSGLMM model outperforming both the MO-sSGLMM and CV-nonSpatial models across different slices in terms of detecting true positives (TP). It should be noted, however, that the MO-sSGLMM model achieves a 100\% precision (no false positives, FP) for most slices, albeit at the cost of a low recall rate (high false negatives, FN), indicating that the model is more conservative in identifying activated voxels. For the CV-nonSpatial model, although it exhibits good precision across the slices, the recall rates remain lower, specifically in the slices with weaker activation strengths (slices 2 and 6). This performance pattern suggests that the model struggles to detect activations in areas with low CNR, highlighting a limitation when dealing with real-world fMRI datasets that often feature low CNR. In comparison, the CV-sSGLMM model consistently detects a higher number of true positives across all slices, demonstrating a stronger detection power even in slices with weak activations (slices 2 and 6). This underscores the benefit of incorporating spatial information, which enhances the model's capacity to detect weaker activations in the presence of complex noise conditions. The model also maintains a 100\% precision across all slices, suggesting that the inclusion of spatial information does not lead to an increase in false positives. As anticipated, both the MO-sSGLMM and CV-sSGLMM models, which employ brain parcellation, demonstrate superior computational efficiency, even when the parallel computation is gated by a 16-core CPU. This advantage becomes even more pronounced when handling larger datasets.

\begin{table}
    \begin{center}
    \renewcommand{\arraystretch}{1.2}
    {\small
    \setlength{\tabcolsep}{3pt}
    \begin{tabular}{c|c|ccccccccc}
    Slice & Model & TP & FP & FN & TN & Precision & Recall & Time (s)\\
    \hline
    \multirow{3}{*}{2} & MO-sSGLMM & 1 & 0 & 49 & 9166 & 1 & 0.02 & \textbf{11.59}\\
     & CV-nonSpatial & 5 & 0 & 45 & 9166 & 1 & 0.1 & 311.55\\
     & CV-sSGLMM & \textbf{8} & 0 & \textbf{42} & 9166 & 1 & \textbf{0.16} & 27.86\\
    \hline
    \multirow{3}{*}{3} & MO-sSGLMM & 22 & 2 & 28 & 9164 & 0.9166 & 0.44 & \multirow{12}{*}{\parbox{1.2cm}{\raggedright same as\newline Slice 2}}\\
     & CV-nonSpatial & 25 & 1 & 25 & 9165 & 0.9615 & 0.50 & \\
     & CV-sSGLMM & \textbf{27} & \textbf{0} & \textbf{23} & \textbf{9166} & \textbf{1} & \textbf{0.54} & \\
    \cline{1-8}
    \multirow{3}{*}{4} & MO-sSGLMM & 30 & 1 & 20 & 9165 & 0.9677 & 0.60 & \\
     & CV-nonSpatial & 30 & 1 & 20 & 9165 & 0.9677 & 060 & \\
     & CV-sSGLMM & \textbf{35} & \textbf{0} & \textbf{15} & \textbf{9166} & \textbf{1} & \textbf{0.70} & \\
    \cline{1-8}
    \multirow{3}{*}{5} & MO-sSGLMM &16 & 5 & 34 & 9161 & 0.7619 & 0.32 & \\
     & CV-nonSpatial & 25 & \textbf{1} & 25 & \textbf{9165} & 0.9615 & 0.50 & \\
     & CV-sSGLMM & \textbf{28} & \textbf{1} & \textbf{22} & \textbf{9165} & \textbf{0.9655} & \textbf{0.56} & \\
    \cline{1-8}
    \multirow{3}{*}{6} & MO-sSGLMM & 0 & 0 & 50 & 9166 & NA & 0 & \\
     & CV-nonSpatial & 4 & 0 & 46 & 9166 & 1 & 0.08 & \\
     & CV-sSGLMM & \textbf{13} & 0 & \textbf{37} & 9166 & \textbf{1} & \textbf{0.26} & \\
    \end{tabular}
    }
    \end{center}
    \caption{Metrics of slices (50 positives and 9166 negatives on each slice) produced by the MO-sSGLMM, CV-nonSpatial, and CV-sSGLMM models}
    \label{tab:SimReal}
\end{table}


\section{Analysis of human CV-fMRI data}\label{RealStudy}
\begin{figure}
    \begin{center}
    \includegraphics[width=\textwidth]{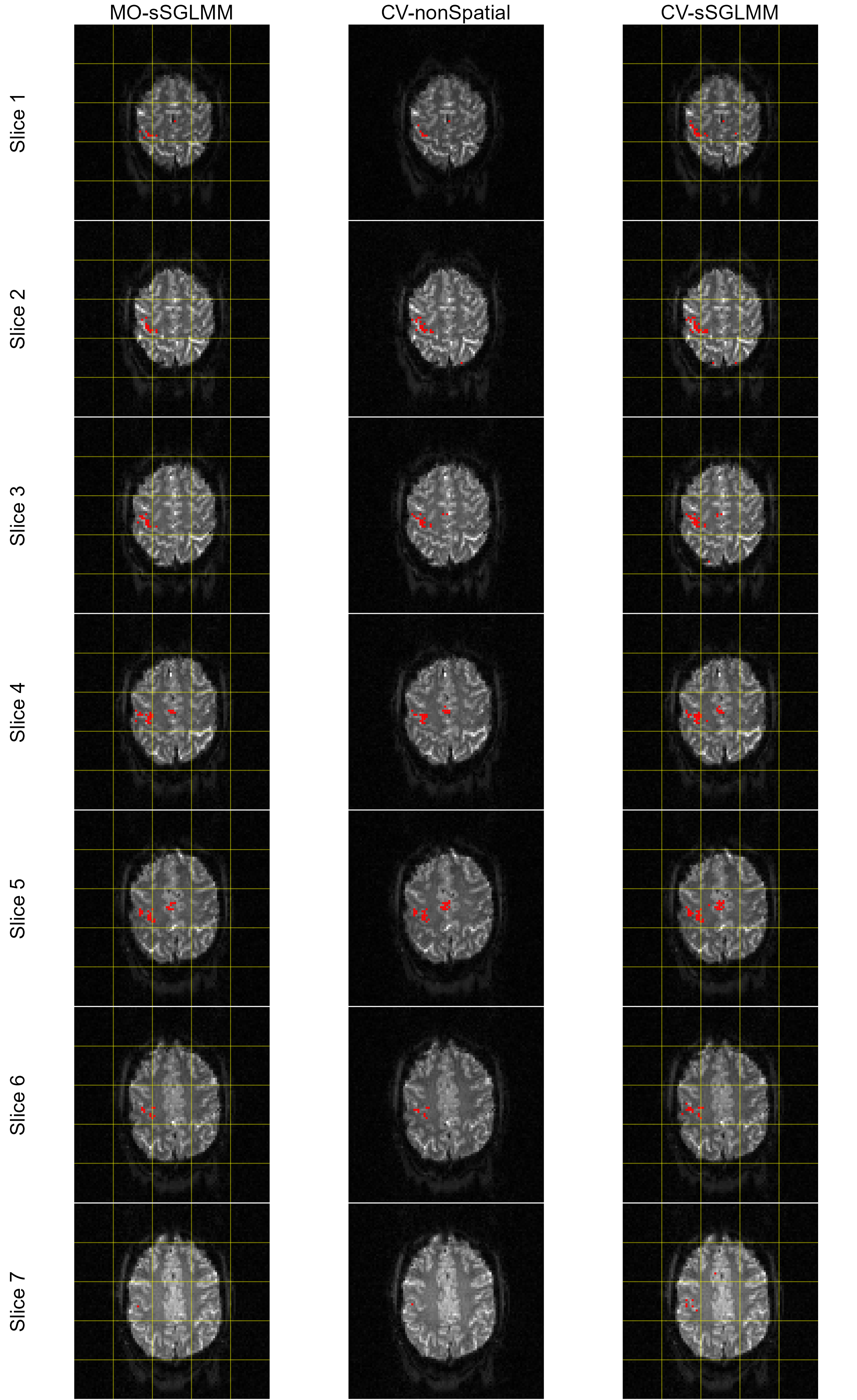}
    \end{center}
    \caption{Estimated activation maps for a real human brain dataset as produced by the MO-sSGLMM, CV-nonSpatial, and CV-sSGLMM models}
    \label{fig:RealHuman_activation_maps}
\end{figure}

\begin{figure}
    \begin{center}
    \includegraphics[width=\textwidth]{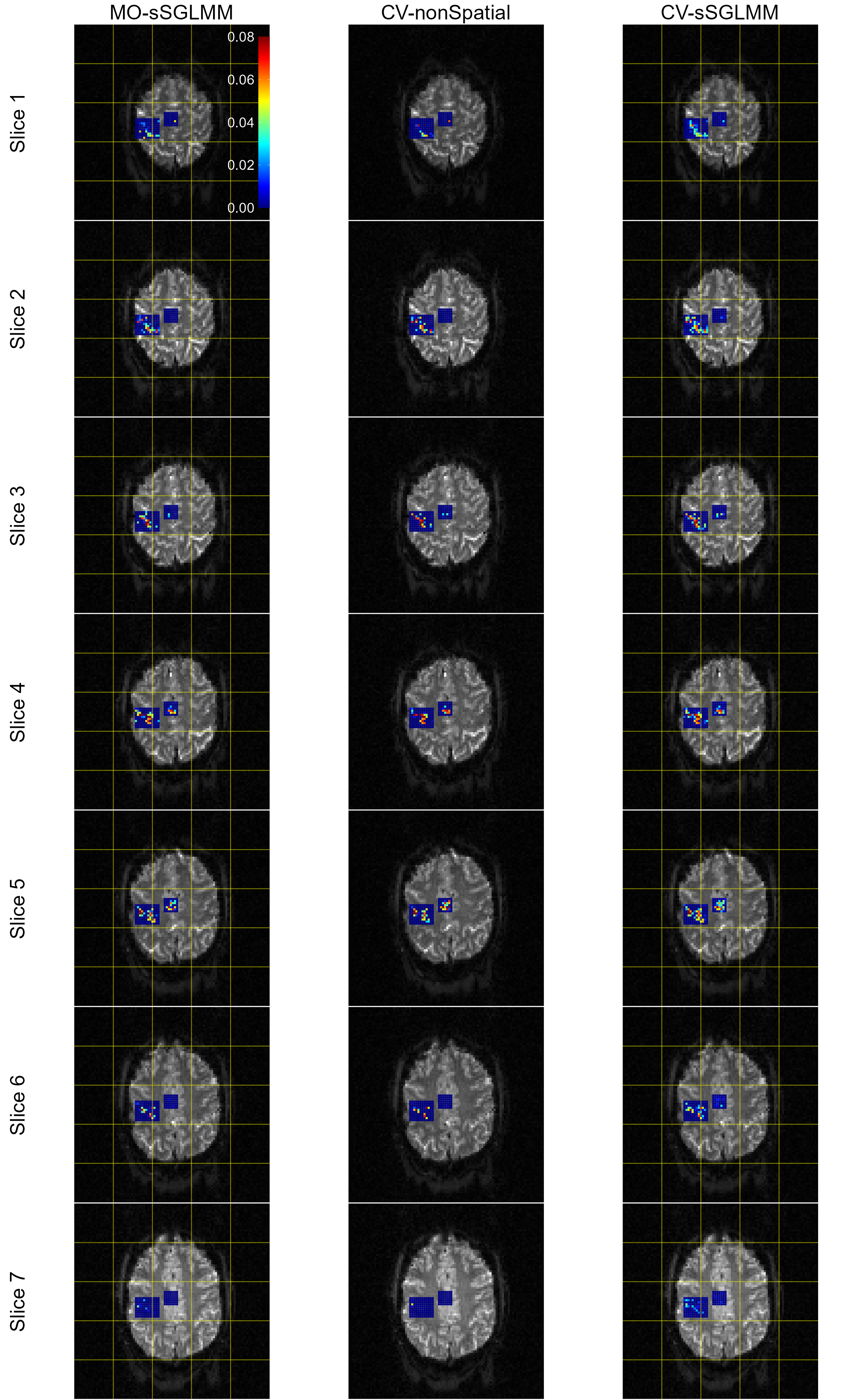}
    \end{center}
    \caption{Estimated magnitude maps for a real human brain dataset as produced by the MO-sSGLMM, CV-nonSpatial, and CV-sSGLMM models}
    \label{fig:RealHuman_strength_maps}
\end{figure}

In this study, we consider the fMRI dataset that is analyzed by \cite{Yu2018}, which is acquired during a unilateral finger-tapping experiment on a 3.0-T General Electric Signa LX MRI scanner. The experimental paradigm involves 16 epochs of alternating 15s on and 15s off periods, leading to $T=490$ time points, including a warm-up period. The data are sourced from seven slices, each of size $96\times 96$. For the MO-sSGLMM and CV-sSGLMM models, we set the parcel number to $G=25$ and again use a threshold of 0.8722 on the inclusion probabilities. The tuning parameter $\psi$ is set to $\Phi^{-1}(0.02)$ and $\Phi^{-1}(0.1)$, respectively. For CV-nonSpatial, the threshold is set to 0.5 as before. The consequent activation and magnitude maps generated from these analyses are depicted in Figure~\ref{fig:RealHuman_activation_maps} and Figure~\ref{fig:RealHuman_strength_maps}. With computation times closely paralleling those in Section \ref{subsec:SimReal} due to comparable dataset sizes, all three models show the same patterns of activation maps. Our CV-sSGLMM consistently demonstrates superior prediction power, particularly evident in the weakly active areas observed in slices 1 and 7, maintaining its consistent performance as discussed in Section \ref{subsec:SimReal}. The active regions identified through our CV-sSGLMM method align with those reported in \cite{Yu2018}, reinforcing the validity of our results and the efficacy of our proposed approach. More importantly, the active regions correspond to areas of the brain that are known to typically be engaged in finger-tapping tasks, affirming the biological relevance of our findings.

\section{Conclusion}\label{Conclusion}
In this study, we propose an innovative fully Bayesian approach to brain activity mapping using complex-valued fMRI data. The proposed model, which incorporates both the real and imaginary components of the fMRI data, provides a holistic perspective on brain activity mapping, overcoming the limitations of the conventional magnitude-only analysis methods. This model showcases the potential to detect task-related activation with higher accuracy. The adoption of an autoregressive error structure, together with spatial priors, allows us to capture both temporal and spatial correlations in brain activity. Moreover, the employment of brain parcellation and parallel computation significantly enhances the model's computational efficiency. Analyses of both simulated and real fMRI data underscores the benefits of our approach, particularly when temporally-correlated, complex-valued noise is present.

There are still areas for exploration. For instance, while we achieve significant results by assuming the phases are constant, we believe that future Bayesian studies based on the dynamic phase model of \cite{Rowe2005c} should be proposed to account for potential phase variations during brain activity \citep{Petridou2013}. Additionally, our current proposal assumes circular data, that is, $\mC^v=\mzero$ for $\mepsilon^v$ in model \eqref{model_complex}, implying that $\beta^v_{Re}$ and $\beta^v_{Im}$ are independent. It would be prudent to develop a more generalized non-circular model where $\mC^v\neq\mzero$ to account for the possibility of non-circular data.


\section*{Acknowledgement}

Research reported in this publication was supported by the National Institute Of General Medical Sciences of the National Institutes of Health under Award Number P20GM139769 (Xinyi Li), National Science Foundation awards DMS-2210658 (Xinyi Li) and DMS-2210686 (D. Andrew Brown). 
The content is solely the responsibility of the authors and does not necessarily represent the official views of the National Institutes of Health or the National Science Foundation.

\clearpage
\bibliographystyle{plainnat}
\bibliography{sample}

\newpage
\appendix

\begin{center}
\huge Appendix
\end{center}


\section{Demonstrating the equivalence between models using real and imaginary parts, and models using magnitude and phase}

This appendix is influenced by \cite{Rowe2009a}, and seeks to demonstrate that, when there's only one stimulus:
\begin{itemize}    
    \item \cite{Lee2007}'s model is approximately equivalent to \cite{Rowe2005c}'s dynamic phase model when the intercept in the magnitude is absent.
    \item \cite{Lee2007}'s model is fully equivalent to \cite{Rowe2004}'s constant phase model.
\end{itemize}
For the first scenario, assuming no intercept in the magnitude, the $v^{\text{th}}$ voxel's complex-valued fMRI signal can be simulated using \cite{Rowe2005c}'s dynamic phase model as per equation:
\begin{equation}\label{append:sim}
    \begin{split}
        \my_{Re}^v&=D_{Re}^v\mx\beta^v,\\
        \my_{Im}^v&=D_{Im}^v\mx\beta^v,
    \end{split}
\end{equation}
where $\my_{Re}^v$ and $\my_{Im}^v$ are simulated complex-valued fMRI vectors of length $T$, and $\mx$ is the expected BOLD response of length $T$ with $\beta^v$ as the scalar magnitude. The matrices $D_{Re}^v$ and $D_{Im}^v$ are $T\times T$ and diagonal with $\cos{(\theta_0+\theta_1x_t)}$ and $\sin{(\theta_0+\theta_1x_t)}$ as the $t^{\text{th}}$ diagonal element, which represent the dynamic phase. By equating this with the means of the \cite{Lee2007}'s model (without intercept), we have:
\begin{equation}
    \begin{split}
        \mx\beta_{Re}^v&=D_{Re}^v\mx\beta^v,\\
        \mx\beta_{Im}^v&=D_{Im}^v\mx\beta^v,
    \end{split}
\end{equation}
where $\beta_{Re}^v$ and $\beta_{Im}^v$ are the scalar real and imaginary parts of the regression coefficient, and the maximum likelihood estimators of them are:
\begin{equation}
    \begin{split}        \widehat{\beta}_{Re}^v&=\left(\mx'\mx\right)^{-1}\mx'D_{Re}^v\mx\beta^v,\\\widehat{\beta}_{Im}^v&=\left(\mx'\mx\right)^{-1}\mx'D_{Im}^v\mx\beta^v,
    \end{split}
\end{equation}
then,
\begin{equation}
\begin{split}
\widehat{\beta}_{Re}^{v,2}+\widehat{\beta}_{Im}^{v, 2}
&=\left[\left(\mx'\mx\right)^{-1}\mx'D_{Re}^v\mx\beta^v\right]^2+\left[\left(\mx'\mx\right)^{-1}\mx'D_{Im}^v\mx\beta^v\right]^2\\
&=\beta^{v, 2}\left(\mx'\mx\right)^{-2}\left[\left(\mx'D_{Re}^v\mx\right)^2+\left(\mx'D_{Im}^v\mx\right)^2\right]\\
&=\beta^{v, 2}\left(\mx'\mx\right)^{-2}\left[\mx'D_{Re}^v\mx\mx'D_{Re}^v\mx+\mx'D_{Im}^v\mx\mx'D_{Im}^v\mx\right]\\
&=\beta^{v, 2}\left(\mx'\mx\right)^{-2}\left[\mx'\left(D_{Re}^v\mx\mx'D_{Re}^v+D_{Im}^v\mx\mx'D_{Im}^v\right)\mx\right].
\end{split}
\end{equation}
Notice that $D_{Re}^v\mx\mx'D_{Re}^v$ and $D_{Im}^v\mx\mx'D_{Im}^v$ are $T\times T$ symmetric matrices with the following terms as the $(i, j)$th element, respectively:
\begin{equation}
    \begin{split}
        &x_ix_j\cos{(\theta_0+\theta_1x_i)}\cos{(\theta_0+\theta_1x_j)},\\
        &x_ix_j\sin{(\theta_0+\theta_1x_i)}\sin{(\theta_0+\theta_1x_j)}.
    \end{split}
\end{equation}
Using the fact that $\cos{(a)}\cos{(b)}+\sin{(a)}\sin{(b)}=\cos{(a-b)}$, we have:
\begin{equation}
    D_{Re}^v\mx\mx'D_{Re}^v+D_{Im}^v\mx\mx'D_{Im}^v=\;\mx\mx'\odot\mP,
\end{equation}
where $\mP$ is a $T\times T$ symmetric matrix and $\mP_{(i, j)} =\cos{\left(\theta_1\left(x_i-x_j\right)\right)}$, and $\odot$ denotes the point-wise product. It's important to note that in both simulated and real data, $\mP$ closely approximates the all-ones matrix $\mone_{T\times T}$. This is because the difference between $x_i$ and $x_j$ is typically small, even when considering the extreme values. After multiplying this small difference with a small $\theta_1$ and then taking the cosine, the result tends to be very close to 1. Thus,
\begin{equation}
\sqrt{\widehat{\beta}_{Re}^{v, 2}+\widehat{\beta}_{Im}^{v, 2}}\approx\sqrt{\beta^{v, 2}\left(\mx'\mx\right)^{-2}\left[\mx'\left(\mx\mx'\right)\mx\right]}=\beta^v.
\end{equation}
In this case, \cite{Lee2007}'s model can be considered as approximately equivalent to \cite{Rowe2005c}'s dynamic phase model.
For the second scenario, when the phase is constant and the intercept is included in the magnitude, using \cite{Rowe2004}'s constant phase model to simulate the data, we get:
\begin{equation}
\begin{split}
    \my_{Re}^v&=\Lambda_{Re}^v\begin{pmatrix} \mone & \mx \end{pmatrix}
    \begin{pmatrix} \beta_0^v \\ \beta_1^v \end{pmatrix},\\
    \my_{Im}^v&=\Lambda_{Im}^v\begin{pmatrix} \mone & \mx \end{pmatrix}
    \begin{pmatrix} \beta_0^v \\ \beta_1^v \end{pmatrix},
\end{split}
\end{equation}
where $\Lambda_{Re}^v=\cos{(\theta)}\,\mI_{T\times T}$ and $\Lambda_{Im}^v=\sin{(\theta)}\,\mI_{T\times T}$. Upon equating this with the means of the \cite{Lee2007}'s model, we have:
\begin{equation}
    \begin{split}
        \begin{pmatrix} \mone & \mx \end{pmatrix}\begin{pmatrix} \beta_{Re, 0}^v \\ \beta_{Re, 1}^v \end{pmatrix}&=\Lambda_{Re}^v\begin{pmatrix} \mone & \mx \end{pmatrix}\begin{pmatrix} \beta_0^v \\ \beta_1^v \end{pmatrix},\\
        \begin{pmatrix} \mone & \mx \end{pmatrix}\begin{pmatrix} \beta_{Im, 0}^v \\ \beta_{Im, 1}^v \end{pmatrix}&=\Lambda_{Im}^v\begin{pmatrix} \mone & \mx \end{pmatrix}\begin{pmatrix} \beta_0^v \\ \beta_1^v \end{pmatrix}.\\
    \end{split}
\end{equation}
Since $\Lambda_{Re}^v$ and $\Lambda_{Im}^v$ don't contain $\mx$, we can remove the means so that to remove the intercept in the model, which yields:
\begin{equation}\label{mode2}
    \begin{split}
        \mx_{c}\beta_{Re, 1}^v&=\Lambda_{Re}^v\mx_{c}\beta_1^v,\\
        \mx_{c}\beta_{Im, 1}^v&=\Lambda_{Im}^v\mx_{c}\beta_1^v,\\
    \end{split}
\end{equation}
where $\mx_{c}$ is the centered $\mx$. This becomes similar to the previous model:
\begin{equation}
    \begin{split}
        \widehat{\beta}_{Re, 1}^{v, 2}+\widehat{\beta}_{Im, 1}^{v, 2}=\beta_1^{v, 2}\left(\mx_c'\mx_c\right)^{-2}&\left[\mx_c'\left(\mx_c\mx_c'\odot\mP\right)\mx_c\right]\\
        &=\beta_1^{v, 2}\left(\mx_c'\mx_c\right)^{-2}\left[\mx_c'\left(\mx_c\mx_c'\right)\mx_c\right]=\beta_1^{v, 2},
    \end{split}
\end{equation}
as $\mP$ is exactly $\mone_{T\times T}$ now. Consequently, \cite{Lee2007}'s model is found to be equivalent to \cite{Rowe2004}'s constant phase model.


\section{Full conditional posterior distributions in the CV-sSGLMM model for Gibbs sampling}
This appendix gives full conditional posterior distributions of $\gamma_v, \mbeta_r^v, \mrho_r^v, \sigma_v^2, \tau_g^2, \eta_v, \mdelta_g, \kappa_g$ for Gibbs sampling. All derivations will omit the subscript of $g$ (parcel index) from the parcel-level parameters $\tau_g^2$, $\mdelta_g$, and $\kappa_g$, since all parcels run the algorithm identically.


\subsection{Full conditional distribution of $\gamma_v$}

For the voxel $v$ ($v=1, ..., V$):
\begin{equation}
p(\gamma_v=1\mid\my_r^v, \mbeta_r^v, \mrho_r^v, \sigma_v^2, \tau^2, \eta_v)
=\frac{p(\gamma_v=1\mid\eta_v)}{p(\gamma_v=1\mid\eta_v)+\frac{L_0}{L_1}{\cdot}p(\gamma_v=0\mid\eta_v)},
\end{equation}
where
\begin{equation}
    \begin{split}
        L_0&=p(\my_r^v, \mbeta_r^v, \mrho_r^v, \sigma_v^2, \tau^2\mid\gamma_v=0),\\
        L_1&=p(\my_r^v, \mbeta_r^v, \mrho_r^v, \sigma_v^2, \tau^2\mid\gamma_v=1).
    \end{split}
\end{equation}
To determine $L_0$ and $L_1$, which are the joint distributions of $\my_r^v, \mbeta_r^v, \mrho_r^v, \sigma_v^2, \tau^2$ under the condition of $\gamma_v=0$ and $\gamma_v=1$, respectively, we recall the CV-sSGLMM model:
\begin{equation}
\my^v=\mx\beta^v+\mr^v\rho^v+\mepsilon^v,\qquad \mepsilon^v\sim\C\N_T(\mzero, 2\sigma_v^2\mI, \mzero).
\end{equation}
Applying Prais-Winsten transformation (order one backward operator) on $\my^v$ and $\mx$, we have:
\begin{equation}
    \begin{split}
        \my^{v*}&=\my_{now}^v-\rho^v\my_{lag1}^v,\\
        \mx^{v*}&=\mx_{now}-\rho^v\mx_{lag1},
    \end{split}
\end{equation}
where $\my_{now}^v$ and $\my_{lag1}^v$ are vectors containing the last and the first $T-1$ elements in $\my^v$, respectively. The vectors $\mx_{now}$ and $\mx_{lag1}$ are from $\mx$ by the same rule of truncation. Now it becomes a model without autoregressive errors:
\begin{equation}
\my^{v*}=\mx^{v*}\beta^v+\mepsilon^v,\qquad \mepsilon^v\sim\C\N_{T-1}(\mzero, 2\sigma_v^2\mI, \mzero),
\end{equation}
with equivalent real-valued representation:
\begin{equation}
\underbrace{\begin{pmatrix}
\my_{Re}^{v*}\\
\my_{Im}^{v*}
\end{pmatrix}}_{\my_r^{v*}}
=
\underbrace{\begin{pmatrix}
\mx_{Re}^{v*} & -\mx_{Im}^{v*}\\
\mx_{Im}^{v*} & \mx_{Re}^{v*}
\end{pmatrix}}_{\mX_r^{v*}}
\underbrace{\begin{pmatrix}
\beta_{Re}^v\\
\beta_{Im}^v
\end{pmatrix}}_{\mbeta_r^v}
+
\underbrace{\begin{pmatrix}
\mepsilon_{Re}^v\\
\mepsilon_{Im}^v
\end{pmatrix}}_{\mepsilon_r^v}.
\end{equation}
Using the symbols in underbraces for a more compact form:
\begin{equation}
\my_r^{v*}=\mX_r^{v*}\mbeta_r^{v}+\mepsilon_r^v, \qquad\mepsilon_r^v\sim\N_{2(T-1)}(\mzero, \sigma_v^2\mI).
\end{equation}
Therefore, when $\gamma_v=1$:
\begin{equation}
    L_1=p(\my_r^v, \mbeta_r^v, \mrho_r^v, \sigma_v^2, \tau^2)\propto p(\my_r^v\mid \mbeta_r^v, \mrho_r^v, \sigma_v^2)\,p(\mbeta_r^v\mid \tau^2),
\end{equation}
where
\begin{equation}
    \begin{split}
        p(\my_r^v\mid\mbeta_r^v, \mrho_r^v, \sigma_v^2)&=(2\pi\sigma_v^2)^{-\frac{2(T-1)}{2}}{\rm exp}\left\{-\frac{1}{2\sigma_v^2}(\my_r^{v*}-\mX_r^{v*}\mbeta_r^{v}){'}(\my_r^{v*}-\mX_r^{v*}\mbeta_r^{v})\right\},\\
        p(\mbeta_r^v\mid{\tau^2})&=(2\pi\tau^2)^{-\frac{2}{2}}{\rm exp}\left\{-\frac{1}{2\tau^2}(\mbeta_r^{v}){'}(\mbeta_r^{v})\right\}.
    \end{split}
\end{equation}
Similarly, when $\gamma_v=0$:
\begin{equation}
    L_0=p(\my_r^v, \mbeta_r^v=\mzero, \mrho_r^v, \sigma_v^2, \tau^2)\propto p(\my_r^v\mid \mbeta_r^v=\mzero, \mrho_r^v, \sigma_v^2)\,p(\mbeta_r^v=\mzero\mid \tau^2),
\end{equation}
where
\begin{equation}
    \begin{split}
        p(\my_r^v\mid\mbeta_r^v=\mzero, \mrho_r^v, \sigma_v^2)&=(2\pi\sigma_v^2)^{-\frac{2(T-1)}{2}}{\rm exp}\left\{-\frac{1}{2\sigma_v^2}(\my_r^{v*}){'}(\my_r^{v*})\right\},\\
        p(\mbeta_r^v=\mzero\mid{\tau^2})&=1.
    \end{split}
\end{equation}
Integrating $\mbeta_r^v$ out of $L_1$ yields:
\begin{equation}
    \begin{split}
        L_1^*
        =&\quad(2\pi\sigma_v^2)^{-\frac{2(T-1)}{2}}\cdot\frac{\sigma_v^2}{\tau^2}\cdot{\rm exp}\left\{-\frac{1}{2\sigma_v^2}(\my_r^{v*}){'}\my_r^{v*}\right\}\left\{{\rm det}\left[(\mX_r^{v*}){'}\mX_r^{v*}+\frac{\sigma_v^2}{\tau^2}\mI\right]\right\}^{-\frac{1}{2}}\\
        &\cdot{\rm exp}\left\{\frac{1}{2\sigma_v^2}\left[(\mX_r^{v*}){'}\my_r^{v*}\right]{'}\left[(\mX_r^{v*}){'}\mX_r^{v*}+\frac{\sigma_v^2}{\tau^2}\mI\right]^{-1}\left[(\mX_r^{v*}){'}\my_r^{v*}\right]\right\}.
    \end{split}
\end{equation}
Then, the ratio is:
\begin{equation}
\frac{L_0}{L_1^*}
        =\frac{\tau^2}{\sigma_v^2}\;\frac{\left\{{\rm det}\left[(\mX_r^{v*}){'}\mX_r^{v*}+\frac{\sigma_v^2}{\tau^2}\mI\right]\right\}^{\frac{1}{2}}}{{\rm exp}\left\{\frac{1}{2\sigma_v^2}\left[(\mX_r^{v*}){'}\my_r^{v*}\right]{'}\left[(\mX_r^{v*}){'}\mX_r^{v*}+\frac{\sigma_v^2}{\tau^2}\mI\right]^{-1}\left[(\mX_r^{v*}){'}\my_r^{v*}\right]\right\}}.
\end{equation}
Using this ratio and $p(\gamma_v=1\mid\eta_v)=\mPhi(\psi+\eta_v)$, the full conditional distribution of $\gamma_v$ is:
\begin{equation}
    \pi(\gamma_v\mid\my_r^v, \mbeta_r^v, \mrho_r^v, \sigma_v^2, \tau^2, \eta_v)={\B}ern\left(P\right),
\end{equation}
where
\begin{equation}
    P=p(\gamma_v=1\mid\my_r^v, \mbeta_r^v, \mrho_r^v, \sigma_v^2, \tau^2, \eta_v)
        =\frac{\mPhi(\psi+\eta_v)}{\mPhi(\psi+\eta_v)+\frac{L_0}{L_1^*}{\cdot}\left[1-\mPhi(\psi+\eta_v)\right]}.
\end{equation}


\subsection{Full conditional distribution of $\mbeta_r^v$}

For the voxels with $\gamma_v=0$, we assign them $\mbeta_r^v=\mzero$. For the voxels with $\gamma_v=1$:
\begin{equation}
    \begin{split}
        \pi(\mbeta_r^v\mid\my_r^v, \mrho_r^v, \sigma_v^2, \tau^2)
        &\propto{p}(\my_r^v\mid\mbeta_r^v, \mrho_r^v, \sigma_v^2)p(\mbeta_r^v\mid  \tau^2)\\
        &\propto{\rm exp}\left\{-\frac{1}{2\sigma_v^2}(\my_r^{v*}-\mX_r^{v*}\mbeta_r^{v}){'}(\my_r^{v*}-\mX_r^{v*}\mbeta_r^{v})\right\}{\rm exp}\left\{-\frac{1}{2\tau^2}(\mbeta_r^{v}){'}(\mbeta_r^{v})\right\}\\
        &\propto{\rm exp}\left\{-\frac{1}{2}\left[(\mbeta_r^v){'}\frac{(\mX_r^{v*}){'}\mX_r^{v*}}{\sigma_v^2}\mbeta_r^v-2(\mbeta_r^v){'}\frac{(\mX_r^{v*}){'}}{\sigma_v^2}\my_r^{v*}+(\mbeta_r^{v}){'}\frac{1}{\tau^2}(\mbeta_r^{v})\right]\right\}\\
        &={\rm exp}\left\{-\frac{1}{2}\left[(\mbeta_r^v){'}\frac{\tau^2(\mX_r^{v*}){'}\mX_r^{v*}+\sigma_v^2\mI}{\sigma_v^2\tau^2}\mbeta_r^v-2(\mbeta_r^v){'}\frac{(\mX_r^{v*}){'}}{\sigma_v^2}\my_r^{v*}\right]\right\},
    \end{split}
\end{equation}
which is a kernel of multivariate normal distribution. Thus:
\begin{equation}
\pi(\mbeta_r^v\mid\my_r^v, \mrho_r^v, \sigma_v^2, \tau^2, \gamma_v=1)=\N_2(\mmu_{\mbeta_r^v}, \mSigma_{\mbeta_r^v}),
\end{equation}
where
\begin{equation}
    \begin{split}
        \mmu_{\mbeta_r^v}&=\left[\frac{\tau^2(\mX_r^{v*}){'}\mX_r^{v*}+\sigma_v^2\mI}{\sigma_v^2\tau^2}\right]^{-1}\frac{(\mX_r^{v*}){'}}{\sigma_v^2}\my_r^{v*}=\left[(\mX_r^{v*}){'}\mX_r^{v*}+\frac{\sigma_v^2}{\tau^2}\mI\right]^{-1}(\mX_r^{v*}){'}\my_r^{v*},\\
        \mSigma_{\mbeta_r^v}&=\left[\frac{\tau^2(\mX_r^{v*}){'}\mX_r^{v*}+\sigma_v^2\mI}{\sigma_v^2\tau^2}\right]^{-1}=\sigma_v^2\left[(\mX_r^{v*}){'}\mX_r^{v*}+\frac{\sigma_v^2}{\tau^2}\mI\right]^{-1}.
    \end{split}
\end{equation}
\\\\{\textbf{Full conditional distribution of $\mrho_r^v$}}\\\\
Since $\mrho_r^v$ is the autoregression coefficient for AR(1) errors, let:
\begin{equation}
\mw^v=\my^v-\mx\beta^v
\end{equation}
be the predicted errors. Let $\mw_{now}^v$ and $\mw_{lag1}^v$ be the vectors containing the last and the first $T-1$ components in $\mw^v$, then:
\begin{equation}
\mw_{now}^v=\mw_{lag1}^v\rho^v+\mepsilon^v, \qquad\mepsilon^v\sim\C\N_{T-1}(\mzero, 2\sigma_v^2\mI, \mzero),
\end{equation}
with equivalent real-valued representation:
\begin{equation}
\underbrace{\begin{pmatrix}
\mw_{now, Re}^{v}\\
\mw_{now, Im}^{v}
\end{pmatrix}}_{\mw_{now, r}^{v}}
=
\underbrace{\begin{pmatrix}
\mw_{lag1, Re}^{v} & -\mw_{lag1, Im}^{v}\\
\mw_{lag1, Im}^{v} & \mw_{lag1, Re}^{v}
\end{pmatrix}}_{\mW_{lag1, r}^{v}}
\underbrace{\begin{pmatrix}
\rho_{Re}^v\\
\rho_{Im}^v
\end{pmatrix}}_{\mrho_r^v}
+
\underbrace{\begin{pmatrix}
\mepsilon_{Re}^v\\
\mepsilon_{Im}^v
\end{pmatrix}}_{\mepsilon_r^v}.
\end{equation}
Using the symbols in underbraces for a more compact form:
\begin{equation}
\mw_{now, r}^{v}=\mW_{lag1, r}^{v}\mrho_r^v+\mepsilon^v, \qquad\mepsilon^v\sim\N_{2(T-1)}(\mzero, \sigma_v^2\mI).
\end{equation}
Assigning a uniform prior, $p(\mrho_r^v)\propto 1$, the full conditional distribution of $\mrho_r^v$ is:
\begin{equation}\label{postrho5}
\pi(\mrho_r^v\mid\my_r^v, \cdot)=\N_2(\mmu_{\mrho_r^v}, \mSigma_{\mrho_r^v}),
\end{equation}
where
\begin{equation}
    \begin{split}
        \mmu_{\mrho_r^v}&=\left[(\mW_{lag1, r}^{v}){'}\mW_{lag1, r}^{v}\right]^{-1}(\mW_{lag1, r}^{v}){'}\mw_{now, r}^{v},\\
        \mSigma_{\mrho_r^v}&=\sigma_v^2\left[(\mW_{lag1, r}^{v}){'}\mW_{lag1, r}^{v}\right]^{-1}.
    \end{split}
\end{equation}


\subsection{Full conditional distribution of $\sigma_r^v$}

The full conditional distribution of $\sigma_r^v$ is also from:
\begin{equation*}
\mw_{now, r}^{v}=\mW_{lag1, r}^{v}\mrho_r^v+\mepsilon^v, \qquad\mepsilon^v\sim\N_{2(T-1)}(\mzero, \sigma_v^2\mI).
\end{equation*}
Assigning a Jeffreys prior, $p(\sigma_v^2)\propto 1/\sigma_v^2$, we have:
\begin{equation}
\pi(\sigma_v^2\mid\my_r^v, \cdot)=\I\G\left(\frac{2(T-1)}{2}, \quad\frac{1}{2}(\mw_{now, r}^{v}-\mW_{lag1, r}^{v}\mrho_r^v){'}(\mw_{now, r}^{v}-\mW_{lag1, r}^{v}\mrho_r^v)\right).
\end{equation}


\subsection{Full conditional distribution of $\tau^2$}

The full conditional distribution of $\tau^2$ should be related to the number of active voxels and could be imposed a Jeffreys prior, $p(\tau^2)\propto 1/\tau^2$. After updating $\mgamma=(\gamma_1, \hdots, \gamma_V)'$ and filtering $\mbeta_r=(\beta_{Re}^1,\cdots,\beta_{Re}^V,\beta_{Im}^1,\cdots,\beta_{Im}^V)'$ by $\mgamma$ to make them strictly zeros and non-zeros in each iteration, we have:
\begin{equation}
\pi(\tau^2\mid\mbeta_r)=\I\G\left(\frac{2\mgamma'\mgamma}{2}, \frac{1}{2}\mbeta_r{'}\mbeta_r\right).
\end{equation}


\subsection{Full conditional distribution of $\eta_v$}

Without considering the condition of $\gamma_v$, we focus on $\pi(\eta_v\mid\kappa)$ first. Let $\mQ_s=\mM{'}\mQ\mM$ and $\mQ_{\kappa s}=\kappa\mQ_s=\kappa\mM{'}\mQ\mM$, then:
\begin{equation}
    \begin{split}
        \pi(\eta_v\mid\kappa)&=\int\pi(\eta_v, \mdelta\mid\kappa)d\mdelta\\
        &=\int\pi(\eta_v\mid\mdelta)\pi(\mdelta\mid\kappa)d\mdelta\\
        &=\int\N(\mm_v{'}\mdelta, 1)\times \N\left(\mzero, \mQ_{\kappa s}^{-1}\right)d\mdelta\\
        &\propto\int{\rm exp}\left\{-\frac{\eta_v^2-2\mm_v{'}\mdelta\eta_v+\mdelta{'}\mm_v\mm_v{'}\mdelta}{2}\right\}{\rm exp}\left\{-\frac{\mdelta{'}\mQ_{\kappa s}\mdelta}{2}\right\}d\mdelta\\
        &={\rm exp}\left\{-\frac{\eta_v^2}{2}\right\}\int{\rm exp}\left\{-\frac{1}{2}\left[\mdelta{'}(\mQ_{\kappa s}+\mm_v\mm_v{'})\mdelta-2\mm_v{'}\mdelta\eta_v\right]\right\}d\mdelta\\
        &={\rm exp}\left\{-\frac{\eta_v^2}{2\left[1-\mm_v{'}(\mQ_{\kappa s}+\mm_v\mm_v{'})^{-1}\mm_v\right]^{-1}}\right\}.
    \end{split}
\end{equation}
Thus, $\eta_v\mid\kappa$ follows normal distribution with mean 0 and variance:
\begin{equation}
\left[1-\mm_v{'}(\mQ_{\kappa s}+\mm_v\mm_v{'})^{-1}\mm_v\right]^{-1}.
\end{equation}
By Woodbury's matrix identity:
\begin{equation}
\left[1-\mm_v{'}(\mQ_{\kappa s}+\mm_v\mm_v{'})^{-1}\mm_v\right]^{-1}=1+\mm_v{'}\mQ_{\kappa s}^{-1}\mm_v.
\end{equation}
That is:
\begin{equation}
\pi(\eta_v\mid\kappa)=\N(0, \quad 1+\mm_v{'}\mQ_{\kappa s}^{-1}\mm_v).
\end{equation}
If the condition of $\gamma_v$ is considered, by \cite{Albert1993}:
\begin{equation}
\pi(\eta_v\mid\gamma_v, \mdelta)= \begin{cases} \T\N(\mm_v{'}\mdelta,\; 1,\; 0,\; \infty) & { {\rm if }\; \gamma_v=1}\\ \T\N(\mm_v{'}\mdelta, \; 1, \; -\infty,\;  0) & { {\rm if }\; \gamma_v=0} \end{cases},
\end{equation}
where $\T\N$ denotes the truncated normal distribution. Thus, when $\gamma_v=1$:
\begin{equation}
    \begin{split}
        \pi(\eta_v\mid\gamma_v ,\kappa)
        &=\int\pi(\eta_v, \mdelta\mid\gamma_v, \kappa)d\mdelta\\
        &=\int\pi(\eta_v\mid\gamma_v, \mdelta)\pi(\mdelta\mid\kappa)d\mdelta\\
        &=\int\T\N(\mm_v{'}\mdelta,\; 1,\; 0,\; \infty)\times \N\left(\mzero, \mQ_{\kappa s}^{-1}\right)d\mdelta\\
        &=\T\N\left(0, \; 1+\mm_v{'}\mQ_{\kappa s}^{-1}\mm_v,\; 0, \;\infty\right).
    \end{split}
\end{equation}
Similarly, when $\gamma_v=0$:
\begin{equation}
    \pi(\eta_v\mid\gamma_v ,\kappa)\
        =\T\N\left(0, \; 1+\mm_v{'}\mQ_{\kappa s}^{-1}\mm_v, \; -\infty, \; 0\right).
\end{equation}
Notice that the variance $1+\mm_v{'}\mQ_{\kappa s}^{-1}\mm_v=1+\mm_v{'}(\kappa\mQ_{s})^{-1}\mm_v$. As $\kappa$ functions as a spatial smoothing parameter, it can be moved out of the parentheses to control the entire variance and play the same role. That is:
\begin{equation}
\pi(\eta_v\mid\gamma_v ,\kappa)= \begin{cases} \T\N(0, \; \frac{1}{\kappa}\underbrace{\left(1+\mm_v{'}\mQ_{s}^{-1}\mm_v\right)}_{\nu_v^2},\; 0,\; \infty) & { {\rm if }\; \gamma_v=1}\\ \T\N(0, \; \frac{1}{\kappa}\underbrace{\left(1+\mm_v{'}\mQ_{s}^{-1}\mm_v\right)}_{\nu_v^2},\; -\infty,\; 0) & { {\rm if }\; \gamma_v=0} \end{cases}.
\end{equation}
Since $\mI+\mM\mQ_{s}^{-1}\mM{'}$ doesn't contain any parameters, it can be pre-calculated, then $\nu_v^2=1+\mm_v{'}\mQ_{s}^{-1}\mm_v$ is its $v^{\text{th}}$ diagonal element. This will accelerate the computation.


\subsection{Full conditional distribution of $\mdelta$}

The full conditional distribution of $\mdelta$ is:
\begin{equation}
    \pi(\mdelta\mid\meta, \kappa)=\N\left(\left(\mQ_{\kappa s}+\mM{'}\mM\right)^{-1}\mM{'}\meta, \quad \left(\mQ_{\kappa s}+\mM{'}\mM\right)^{-1}\right).
\end{equation}
Similar to how we deal with $\kappa$ for $\eta_v$, this distribution becomes:
\begin{equation}
\pi(\mdelta\mid\meta, \kappa)=\N\left(\frac{1}{\kappa}\underbrace{\left(\mQ_{s}+\mM{'}\mM\right)^{-1}}_{\widehat{\mQ}_s^{-1}}\mM{'}\meta, \quad \frac{1}{\kappa}\underbrace{\left(\mQ_{s}+\mM{'}\mM\right)^{-1}}_{\widehat{\mQ}_s^{-1}}\right),
\end{equation}
where $\widehat{\mQ}_s^{-1}=\left(\mQ_{s}+\mM{'}\mM\right)^{-1}$ can be pre-calculated to accelerate the computation.


\subsection{Full conditional distribution of $\kappa$}

We assume $\eta_1, ..., \eta_V$ are conditionally independent when given $\kappa$, thus:
\begin{equation}
    \begin{split}
        \pi(\meta\mid\kappa)
        =&\prod_{v=1}^{V}\pi(\eta_v\mid\kappa)\\
        =&\left[\left(\frac{1}{\kappa}\right)^{-\frac{V}{2}}\prod_{v=1}^{V}\left(1+\mm_v{'}\mQ_{s}^{-1}\mm_v\right)^{-\frac{1}{2}}\right]{\rm exp}\left\{-\sum_{v=1}^{V}\frac{\eta_v^2}{2\cdot\frac{1}{\kappa}\cdot(1+\mm_v{'}\mQ_{s}^{-1}\mm_v)}\right\}\\
        \propto&\;\kappa^{\frac{V}{2}}\cdot{\rm exp}\left\{-\kappa\cdot\frac{1}{2}\cdot\sum_{v=1}^{V}\frac{\eta_v^2}{(1+\mm_v{'}\mQ_{s}^{-1}\mm_v)}\right\}.
    \end{split}
\end{equation}
Therefore, the full conditional distribution of $\kappa$ is:
\begin{equation}\label{postkappa4}
    \begin{split}
        \pi(\kappa\mid\meta)&\propto\pi(\meta\mid\kappa)\pi(\kappa)\\
        &\propto\kappa^{\frac{V}{2}}\cdot{\rm exp}\left\{-\kappa\cdot\frac{1}{2}\cdot\sum_{v=1}^{V}\frac{\eta_v^2}{(1+\mm_v{'}\mQ_{s}^{-1}\mm_v)}\right\}\cdot\kappa^{\frac{1}{2}-1}\cdot{\rm exp}\left\{-\frac{\kappa}{2000}\right\}\\
        &=\kappa^{\frac{V+1}{2}-1}{\rm exp}\left\{-\kappa\left[\frac{1}{2}\left(\sum_{v=1}^{V}\frac{\eta_v^2}{(1+\mm_v{'}\mQ_{s}^{-1}\mm_v)}\right)+\frac{1}{2000}\right]\right\}.
    \end{split}
\end{equation}
That is:
\begin{equation}
    \begin{split}
        \pi(\kappa\mid\meta)
        &=\G amma\left(a=\frac{V+1}{2},\quad b=\left[\frac{1}{2}\left(\sum_{v=1}^{V}\frac{\eta_v^2}{(1+\mm_v{'}\mQ_{s}^{-1}\mm_v)}\right)+\frac{1}{2000}\right]^{-1}\right)\\
        &=\G amma\left(a=\frac{V+1}{2},\quad b=\left[\frac{1}{2}\left(\frac{\eta_1^2}{\nu_1^2}+\cdots+\frac{\eta_V^2}{\nu_V^2}\right)+\frac{1}{2000}\right]^{-1}\right),
    \end{split}
\end{equation}
where $b$ is the scale, and the details for $\nu_v^2$ are in the full conditional distribution of $\eta_v$.


\section{More estimations by the CV-sSGLMM model}

\begin{figure}
\begin{center}
\includegraphics[width=\textwidth]{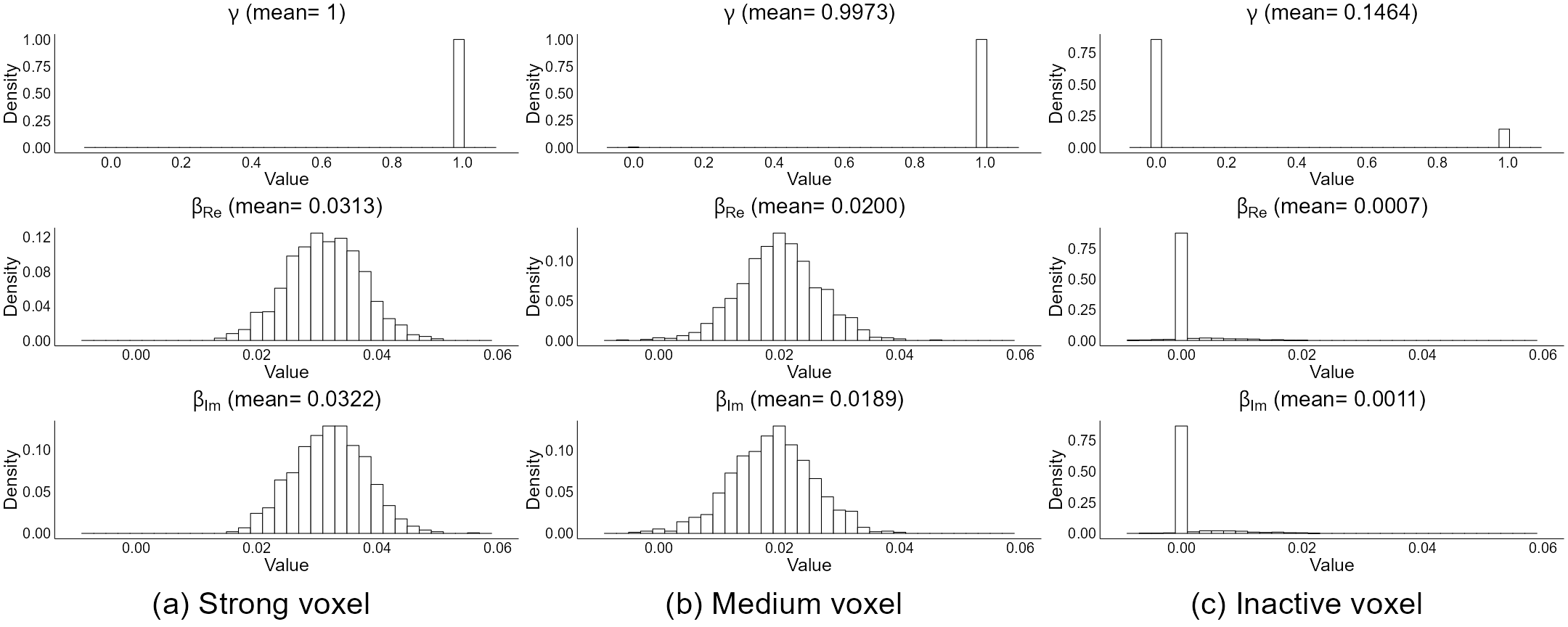}
\end{center}
\caption{(a) is marginal posterior distributions of $\gamma$, $\beta_{Re}$, and $\beta_{Im}$ for a voxel exhibiting high magnitude. (b)-(c) are similar distributions for a medium-magnitude voxel and an inactive voxel, respectively.}
\label{fig:Histogram}
\end{figure}

\begin{figure}
\begin{center}
\includegraphics[width=\textwidth]{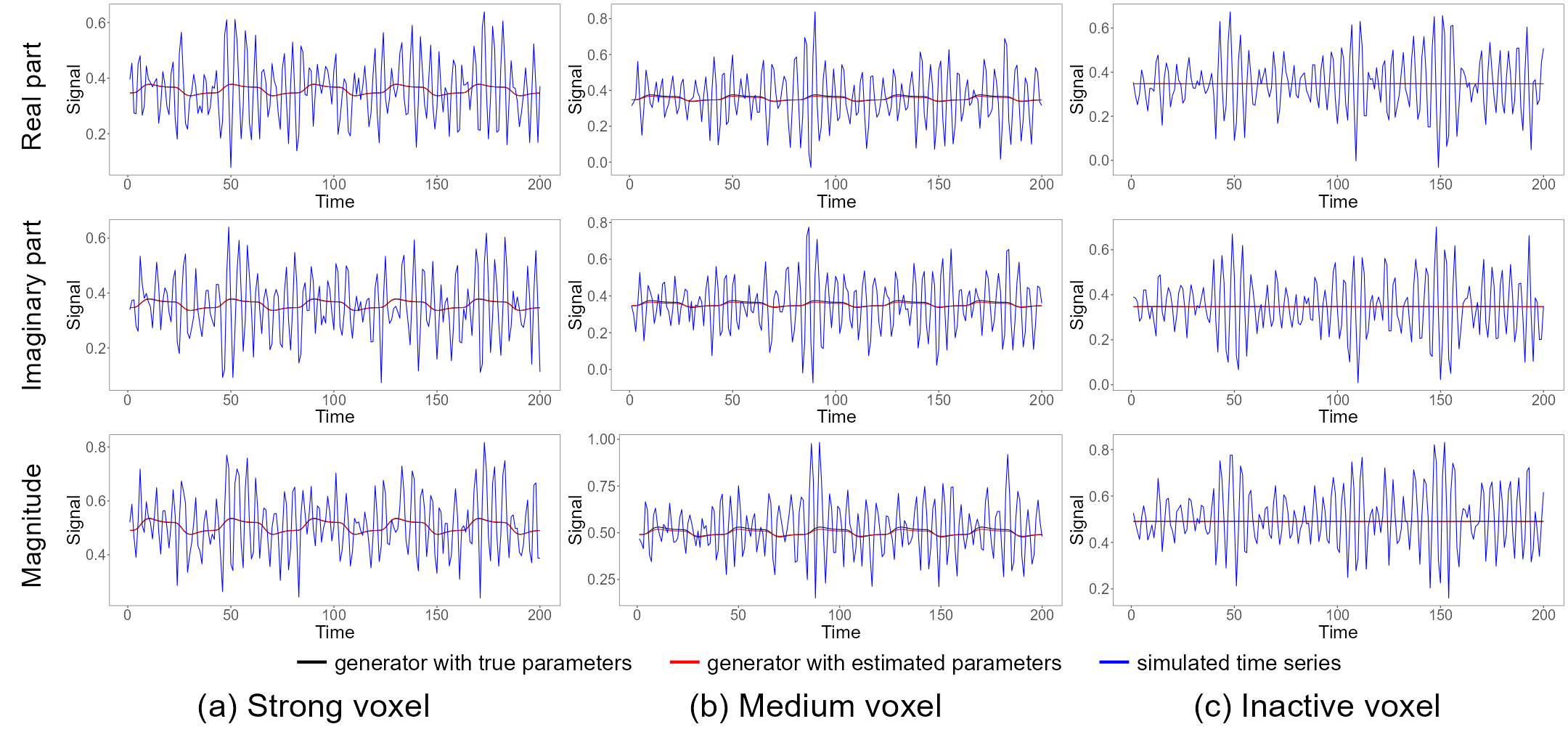}
\end{center}
\caption{(a) is time series of generator using true parameters, generator using estimated parameters, and simulated time series (generator using true parameters with noise) of a voxel exhibiting high magnitude. (b)-(c) are similar time series for a medium-magnitude voxel and an inactive voxel, respectively.}
\label{fig:Time_series}
\end{figure}

\begin{figure}
\begin{center}
\includegraphics[width=\textwidth]{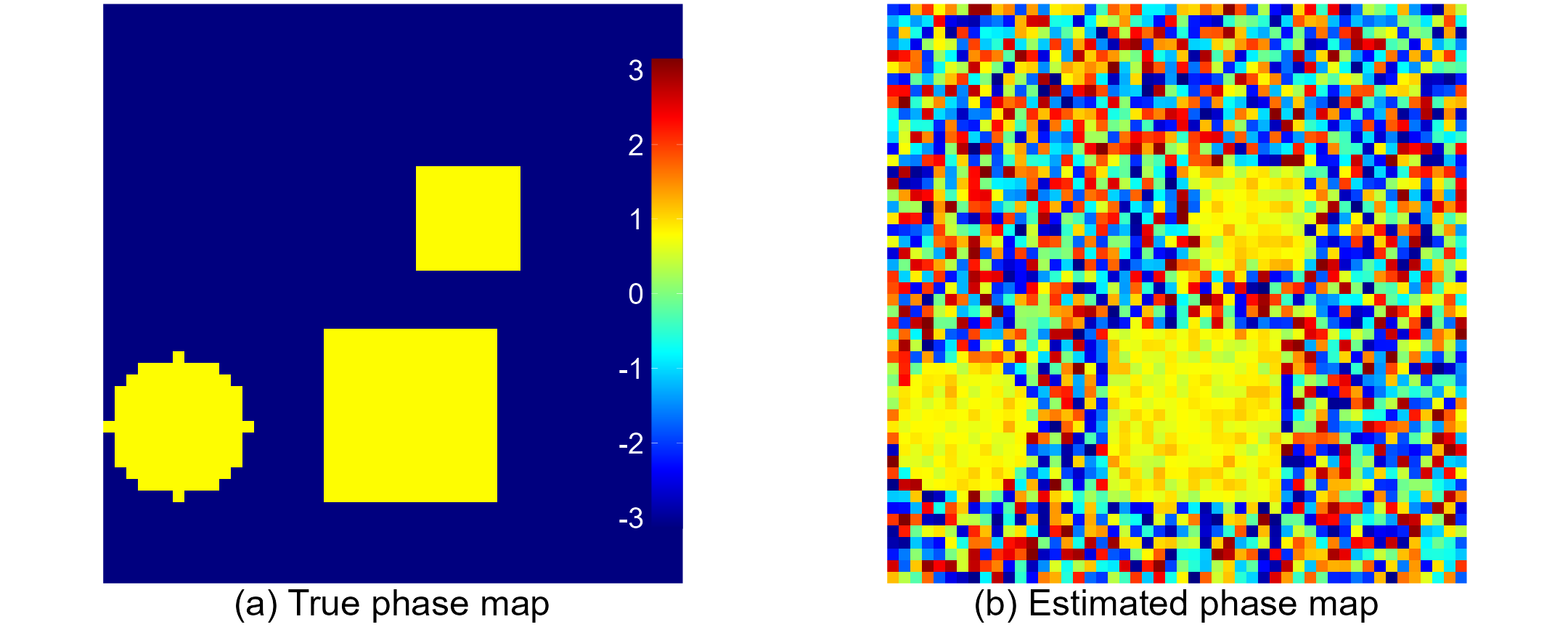}
\end{center}
\caption{(a) is the true phase map of an AR(1) dataset. (b) is the estimated phase map as produced by the CV-sSGLMM model.}
\label{fig:Phase}
\end{figure}

The CV-sSGLMM model is applied to estimate the marginal posterior distributions from three distinct types of voxels (strongly active, moderately active, inactive) within an AR(1) dataset, as showcased in Figure~\ref{fig:Histogram}. The bell-shaped distributions of $\beta_{Re}$ and $\beta_{Im}$ corroborate the theoretical derivation and affirm the reliable performance of the MCMC algorithm during the sampling process. The true and estimated time series from these three voxel are presented in Figure~\ref{fig:Time_series}. The congruence between the generator using true parameters (in black) and that using estimated parameters (in red) is evident. Additionally, both sets of time series aptly capture the pattern of the simulated time series (in blue). This alignment serves as a further testament to the good estimation performance of our CV-sSGLMM model. The phase of voxels is also estimated by the CV-sSGLMM model, and the outcomes are displayed in Figure~\ref{fig:Phase}. Figure~\ref{fig:Phase}(a) presents the true phase map, simulated using a constant phase value of $\theta=\pi/4\approx 0.79$ for active voxels. Figure~\ref{fig:Phase}(b) demonstrates that the CV-sSGLMM model effectively estimated this phase map by $\widehat{\theta}_v=\arctan{\left(\widehat{\beta}_{Im}^v/\widehat{\beta}_{Rm}^v\right)}$.

\end{document}